\documentclass{article}
\usepackage{graphicx}  % standard LaTeX graphics tool;
\usepackage{amsmath}   % For getting proper math eqns
\usepackage[compress]{cite}
\usepackage{amssymb}   % Used for getting various symbols eg. gtrsim, lesssim etc.
\usepackage{bm} % For bold math (esp of lower greek letters)
\usepackage{dcolumn}% Align table columns on decimal point
\usepackage{color}
\usepackage{mathrsfs}
\usepackage{amsfonts}
\usepackage{varioref}
\usepackage{float}
\usepackage[caption=false]{subfig}
\RequirePackage[colorlinks,citecolor=blue,urlcolor=magenta,linkcolor=blue]{hyperref}
\usepackage{tablefootnote}
\def\gr{general relativity}
\def\RN{Reissner-Nordstr\"{o}m }
\def\KN{Kerr-Newman }

\labelformat{section}{Section #1} 
\labelformat{subsection}{Section #1} 
\labelformat{subsubsection}{Section #1}
\labelformat{subsubsubsection}{Section #1}
\labelformat{equation}{Eq.~(#1)} 
\labelformat{figure}{Fig.~#1} 
\labelformat{subfigure}{Fig.~\thefigure#1} 
\labelformat{table}{Table~#1} 
\labelformat{appendix}{Appendix #1}
\title{\bf Looking for extra dimensions in the observed quasi-periodic oscillations of black holes}

\author{Indrani Banerjee\footnote{banerjeein@nitrkl.ac.in}~$^{1}$, Sumanta Chakraborty\footnote{sumantac.physics@gmail.com}~$^{2}$ and Soumitra SenGupta\footnote{tpssg@iacs.res.in}~$^{2}$\\
{\small{$^{1}$Department of Physics and Astronomy, National Institute of Technology, Rourkela-769008, India}}\\
{\small{$^{2}$School of Physical Sciences, Indian Association for the Cultivation of Science, Kolkata-700032, India}}}
\begin{document}
  
\maketitle
%%%%%%%%%%%%%%%%%%%%%%%%%%%%%%%%%%%%%%%%%%%%%%%%%%%%%%%%%%%%%%%%%%%%%%%%%%%%%%%%%%%%%%%%%%%%%%%%%%%
%%%%%%%%%%%%%%%%%%%%%%%%%%%%%%%%%%%%%%%%%%%%%%%%%%%%%%%%%%%%%%%%%%%%%%%%%%%%%%%%%%%%%%%%%%%%%%%%%%%
%%%%%%%%%%%%%%%%%%%%%%%%%%%%%%%%%%%%%%%%%%%%%%%%%%%%%%%%%%%%%%%%%%%%%%%%%%%%%%%%%%%%%%%%%%%%%%%%%%%
\begin{abstract}
Quasi-periodic oscillations, often present in the power density spectrum of accretion disk around black holes, are useful probes for the understanding of gravitational interaction in the near-horizon regime of black holes. Since the presence of an extra spatial dimension modifies the near horizon geometry of black holes, it is expected that the study of these quasi-periodic oscillations may shed some light on the possible existence of these extra dimensions. Intriguingly, most of the extra dimensional models, which are of significant interest to the scientific community, predicts the existence of a tidal charge parameter in black hole spacetimes. This tidal charge parameter can have an overall negative sign and is a distinctive signature of the extra dimensions. Motivated by this, we have studied the quasi-periodic oscillations for a rotating braneworld black hole using the available theoretical models. Subsequently, we have used the observations of the quasi-periodic oscillations from available black hole sources, e.g., GRO J1655 --- 40, XTE J1550 --- 564, GRS 1915 + 105, H 1743 + 322 and Sgr A* and have compared them with the predictions from the relevant theoretical models, in order to estimate the tidal charge parameter. It turns out that among the 11 theoretical models considered here, 8 of them predict a negative value for the tidal charge parameter, while for the others negative values of the tidal charge parameter are also well within the 1-$\sigma$ confidence interval. 
\end{abstract}
%%%%%%%%%%%%%%%%%%%%%%%%%%%%%%%%%%%%%%%%%%%%%%%%%%%%%%%%%%%%%%%%%%%%%%%%%%%%%%%%%%%%%%%%%%%%%%%%%%%
%%%%%%%%%%%%%%%%%%%%%%%%%%%%%%%%%%%%%%%%%%%%%%%%%%%%%%%%%%%%%%%%%%%%%%%%%%%%%%%%%%%%%%%%%%%%%%%%%%%
%%%%%%%%%%%%%%%%%%%%%%%%%%%%%%%%%%%%%%%%%%%%%%%%%%%%%%%%%%%%%%%%%%%%%%%%%%%%%%%%%%%%%%%%%%%%%%%%%%%
%\newpage
%%%%%%%%%%%%%%%%%%%%%%%%%%%%%%%%%%%%%%%%%%%%%%%%%%%%%%%%%%%%%%%%%%%%%%%%%%%%%%%%%%%%%%%%%%%%%%%%%%%
%%%%%%%%%%%%%%%%%%%%%%%%%%%%%%%%%%%%%%%%%%%%%%%%%%%%%%%%%%%%%%%%%%%%%%%%%%%%%%%%%%%%%%%%%%%%%%%%%%%
%%%%%%%%%%%%%%%%%%%%%%%%%%%%%%%%%%%%%%%%%%%%%%%%%%%%%%%%%%%%%%%%%%%%%%%%%%%%%%%%%%%%%%%%%%%%%%%%%%%
\section{Introduction}\label{QPO_Intro}

Discovery of gravitational waves from the merger between black holes and neutron stars \cite{Abbott:2017vtc,TheLIGOScientific:2016pea,Abbott:2016nmj,TheLIGOScientific:2016src,Abbott:2016blz}, along with the detection of the shadow of the supermassive black hole M87* \cite{Fish:2016jil,Akiyama:2019cqa,Akiyama:2019brx,Akiyama:2019sww,Akiyama:2019bqs,Akiyama:2019fyp,Akiyama:2019eap}, have tested the gravitational interaction in the strong gravity regime (aka, the region near the event horizon) to an unprecedented level. So far, both of these, so-called \emph{strong field tests of gravity}, are consistent with the predictions from \gr, which is the most successful theory in explaining the gravitational interaction in the weak field regime \cite{Will:2005yc,Will:1993ns,Will:2005va,Berti:2015itd}. Despite the astounding success of \gr, there exist a few regimes where the theory seems to be breaking down, e.g., near the black hole and cosmological singularities, where quantum effects of gravity become important \cite{Penrose:1964wq,Hawking:1976ra,Wald,Christodoulou:1991yfa}. Moreover, \gr\ also falls short in explaining the dark sector, i.e., dark matter and dark energy, which are invoked to explain the flat rotation curves of galaxies and the accelerated expansion of the universe \cite{Clifton:2011jh,Perlmutter:1998np,Riess:1998cb}. This has resulted into a proliferation of alternate gravity models, which can be strong contenders of \gr, such that they match with \gr\ for weak gravitational field, but differing in the strong field or large scale behaviours. 

Among the various proposals for alternative gravity theories, the prominent ones include --- (a) $f(R)$ gravity \cite{Nojiri:2010wj,Nojiri:2003ft}, Lovelock theories \cite{Lanczos:1932zz,Lovelock:1971yv,Padmanabhan:2013xyr,Dadhich:2015ivt}, scalar-tensor/Horndeski models \cite{Horndeski:1974wa,Sotiriou:2013qea,Babichev:2016rlq} and theories with extra spatial dimensions \cite{Antoniadis:1990ew,ArkaniHamed:1998rs,Randall:1999vf,Randall:1999ee,Shiromizu:1999wj,Dadhich:2000am}. In what follows, we will explore the modifications to \gr\ due to the presence of an extra spatial dimension, since it offers one of the simplest and minimal modification to the gravitational interaction in the four-dimensional spacetime. These extra dimensional models, also known as the brane-world models \cite{Randall:1999vf,Randall:1999ee,Garriga:1999yh,Csaki:1999mp}, are designed to provide a resolution to the fine-tuning problem (or, the gauge hierarchy problem) in particle physics \cite{Antoniadis:1990ew,ArkaniHamed:1998rs,Randall:1999ee,Goldberger:1999uk}. The particular model we will consider in this work, assumes that all the Standard Model particles and fields are confined to the four-dimensional observable universe known as the 3-brane (or, simply as the \emph{brane}), while gravity pervades the full five-dimensional spacetime, known as the \emph{bulk} \cite{Dadhich:2000am,Harko:2004ui,Shiromizu:1999wj,Randall:1999vf}. Even though the brane-world model has appealing theoretical features, it is important to consider experimental/observational implications of this model. As evident, the particle physics experiments often provide strong bounds on the extra dimensional models \cite{Dimopoulos:2001hw,Davoudiasl:1999jd,Davoudiasl:1999tf}, however in the present context the Standard Model particles and fields are all confined to four dimensions and hence the most promising way to test the present brane-world scenario is through gravitational interaction. In the context of gravitational waves, the presence of an extra spatial dimension modifies the black hole quasi-normal modes and hence provides distinct signature of the extra dimension \cite{Berti:2009kk,Kanti:2005xa,Konoplya:2011qq,Toshmatov:2016bsb,Andriot:2017oaz,Chakraborty:2017qve}. In addition, the presence of extra dimensions also affect the equation of state of a neutron star, thereby modifying its deformability, which can be used to impose tight constraints on the extra dimensional parameters, using for example, the GW170817 event \cite{Hinderer:2007mb,Cardoso:2017cfl,TheLIGOScientific:2017qsa,Chakravarti:2018vlt,Chakravarti:2019aup}. Furthermore, the presence of extra dimension also advertise in favour of exotic compact objects, having distinct signature compared to black holes in \gr\ in various astrophysical circumstances \cite{Dey:2020lhq,Dey:2020pth,Chakraborty:2021gdf}. Intriguingly, the footprints of extra dimension can also be found in the measurement of black hole shadow as well. As we have demonstrated in \cite{Banerjee:2019nnj}, the shadow measurement of M87* is favouring the presence of an extra dimension more than \gr. 

All these results motivate us to search for further and possibly more conclusive evidence for extra dimensions in other avenues, e.g., in the quasi-periodic oscillation from black holes. This is what we wish to study in this work. For this purpose, we consider a four dimensional brane depicting the visible universe, embedded in a five dimensional bulk. As a consequence, the non-local effects of the bulk Weyl tensor acts as a source to the four dimensional gravity, modifying the Einstein's equations, even in the absence of any matter-energy on the brane \cite{Shiromizu:1999wj,Harko:2004ui,Dadhich:2000am,Chakraborty:2015bja}. A certain class of vacuum solutions of the modified field equations resemble the \RN metric in \gr\ if the background spacetime is static and spherically symmetric \cite{Harko:2004ui} or the \KN solution in case the metric is stationary and axi-symmetric \cite{Aliev:2005bi}. However, unlike \gr, the tidal charge parameter in the aforesaid spacetimes owes its origin to extra dimensions and hence can also assume negative values, which turns out to be a distinguishing signature of higher dimensions \cite{Schee:2008kz}.

In this work, we focus on the role of the tidal charge parameter, arising in the aforementioned braneworld scenario, in explaining the quasi-periodic oscillations (henceforth as QPOs) observed in the power density spectrum of black holes (henceforth as BHs). QPOs are primarily observed in the power spectrum of several Low-Mass X-ray binaries (henceforth as LMXRBs) \cite{2006csxs.book.....L,vanderKlis:2000ca}, including neutron star (henceforth as NS) and BH sources, although some active galactic nuclei (henceforth as AGNs) also exhibit QPOs in the X-ray flux \cite{2008Natur.455..369G}. The QPOs appear as peaks in the power spectrum of BHs and NSs and these peaks are believed to hold important information about the nature of gravitational interaction in the vicinity of these objects and can potentially serve as an effective tool to test \gr\ and its alternatives \cite{2006csxs.book.....L,vanderKlis:2000ca}. In particular, the observed frequency range of the QPOs varies from mHz to hundreds of Hz for stellar mass BHs, while for NSs the range is from deciHz to KHz order \cite{vanderKlis:2000ca,Maselli:2014fca}. The high frequency QPOs (henceforth as HFQPOs), on the other hand, in these LMXRBs are particularly interesting since they involve timescales $\sim 0.1–1~ \rm ms$, which are close to the dynamical time scales of accreting matter near the vicinity ($r<10~R_{\rm g}$) of these compact objects, where gravity is expected to be the strongest \cite{PhysRevLett8217}. These timescales can be directly inferred from the fact that the characteristic velocities of accreting fluids near these compact objects scale as, $v\sim \sqrt{(GM/r)}$, such that the dynamical timescale becomes, $t_{\rm d}\sim \sqrt{(r^{3}/GM)}$. Following which, one obtains the dynamical timescale to read,  $t_{\rm d}\sim ~0.1~\rm ms$ at $r\sim 15~\textrm{km}$ for a $1.4~M_\odot$ NS, while the dynamical timescale becomes, $t_{\rm d}\sim 1~\textrm{ms}$ at $r\sim 100~\textrm{km}$ for a $10~M_\odot$ BH \cite{2006csxs.book.....L,vanderKlis:2000ca}. The above numerical estimation demonstrates that the dynamical timescales are in the Millisecond range, which have been predicted back in the 1970s \cite{1971SvA....15..377S,1973SvA....16..941S} and achieved observational confirmation with the launch of NASA's Rossi X-Ray Timing Explorer satellite \cite{2006csxs.book.....L}. Since these Millisecond order dynamical timescales arises out of physics close to the event horizon, QPOs provide a unique opportunity to study both the physics of dense matter in neutron star (NS) \cite{vanderKlis:2000ca} as well as the near horizon physics of BHs. However, NSs come with additional complications, e.g., the existence of stable magnetic fields or, the existence of a boundary layer affecting the dynamics of accreting matter in the vicinity of these objects \cite{Maselli:2014fca}. Thus owing to their simpler structure we will explore the observed QPOs in BHs to decipher the signatures of extra dimensions. In particular, our earlier works \cite{Banerjee:2017npv,Banerjee:2019sae,Banerjee:2019cjk,Banerjee:2019nnj} have consistently suggested the possibility of a negative tidal charge parameter through various BH observations. This has further motivated us to observe if a similar conclusion regarding the tidal charge holds true for the QPOs as well, which will enable us to understand if such an observation is consistent with our previous findings.   

The paper is organized as follows: In \ref{S2} we briefly review the brane world model, explaining the modifications in the gravitational field equations due to the presence of extra dimensions. The stationary axisymmetric black hole solution with tidal charge has also been discussed. \ref{S3} is dedicated in deriving the epicyclic frequencies of matter orbiting a given stationary axisymmetric spacetime while the existing QPO models which relate the observed QPO frequencies with these epicyclic frequencies are reviewed in \ref{S4}. A chi-square analysis comparing the theoretical predictions from the various QPO models with the available QPO observations of BHs is presented in \ref{S5}. We conclude with a summary of our results with some scope for future works in \ref{S6}.  

\textit{Notations and Conventions:} Throughout this paper, the Greek indices denote the four dimensional spacetime coordinates and we will work in geometrized unit with $G=1=c$. The metric convention adopted in this work will be mostly positive, i.e., the flat spacetime metric has the form $\textrm{diag}(-,+,+,+)$.  

%%%%%%%%%%%%%%%%%%%%%%%%%%%%%%%%%%%%%%%%%%%%%%%%%%%%%%%%%%%%%%%%%%%%%%%%%%%%%%%%%%%%%%%%%%%%%%%%%%%
%%%%%%%%%%%%%%%%%%%%%%%%%%%%%%%%%%%%%%%%%%%%%%%%%%%%%%%%%%%%%%%%%%%%%%%%%%%%%%%%%%%%%%%%%%%%%%%%%%%
%%%%%%%%%%%%%%%%%%%%%%%%%%%%%%%%%%%%%%%%%%%%%%%%%%%%%%%%%%%%%%%%%%%%%%%%%%%%%%%%%%%%%%%%%%%%%%%%%%%
\section{Basics of rotating black hole in the brane world gravity}\label{S2}
 
In this section we will briefly outline the basic properties of the vacuum, axisymmetric solution to the effective gravitational field equations on the brane, which can be achieved by projecting the bulk gravitational field equations on the brane hypersurface. This route of obtaining the effective gravitational field equations on the brane is convenient, since we can bypass any requirement of a detailed knowledge of the bulk geometry, which a priori is absent to any brane observer. If the bulk gravitational field equations are the bulk Einstein's equations, then the Gauss, Codazzi and Mainardi equations can be used to project the bulk Einstein tensor on the brane hypersurface and thereby arriving at the effective gravitational field equations on the brane \cite{Shiromizu:1999wj,Harko:2004ui,Chakraborty:2014xla,Chakraborty:2015bja}. The effective field equations for vacuum brane involve --- (a) the four-dimensional Einstein tensor $G_{\mu \nu}$, having no connection to the extra spatial dimension and (b) projection of the bulk Weyl tensor $W_{ABCD}$, onto the brane hypersurface, yielding $E_{\mu \nu}=W_{ABCD}e^{A}_{\mu}n^{B}e^{C}_{\nu}n^{D}$, where $e^{A}_{\mu}$ are the projectors and $n_{A}$ are the normals to the brane hypersurface. In the case of non-vacuum brane, there are of course additional contributions from the matter sector, which we will ignore here. It is worth mentioning that the presence of an extra spatial dimension for vacuum brane is contained in the effective gravitational field equations through the term $E_{\mu \nu}$, generated from the bulk Weyl tensor. By virtue of its connection to the Weyl tensor and the symmetries of the same, it follows that the tensor $E_{\mu \nu}$ is traceless and hence mimics the energy-momentum tensor of a Maxwell field with a crucial overall negative sign \cite{Aliev:2005bi}. The static and spherically symmetric vacuum brane spacetime, arising out of the effective gravitational field equations on the brane, has already been investigated in \cite{Dadhich:2000am}. While the stationary and axisymmetric counterpart of the same, which is observationally more relevant \cite{Aliev:2005bi}, will be the prime focus of the present work. 

Since the tensor $E_{\mu \nu}$ has the same properties as that of the energy-momentum tensor of the Maxwell field, the resulting stationary and axi-symmetric solution will look like the Kerr-Newman spacetime, such that the associated line element takes the following form,
%%%%%%%%%%%%%%%%%%%%%%%%%%%%%%%%%%%%%%%%%%%%%%%%%%%%%%%
\begin{align}\label{2-2}
ds^{2}&=-\bigg(1-\frac{2Mr-M^{2}q}{\rho^{2}}\bigg)dt^2-\frac{2a\sin^2\theta(2Mr-M^{2}q)}{\rho^{2}} dt d\phi 
+ \frac{\rho^{2}}{\Delta}dr^2 +\rho^{2} d\theta^2 
\nonumber
\\
&\qquad \qquad+\bigg\{r^2 + a^2 +\frac{a^2 \sin^2\theta\left(2Mr-M^{2}q\right)}{\rho^{2}}\bigg\}\sin^2\theta d\phi^2~.
\end{align}
%%%%%%%%%%%%%%%%%%%%%%%%%%%%%%%%%%%%%%%%%%%%%%%%%%%%%%%%
In the above line element for notational convenience the following shorthand definitions were used: $\rho^{2} \equiv r^2+a^2 \cos^{2}\theta$ and $\Delta \equiv r^{2}-2Mr+a^{2}+M^{2}q$.

Note that the dimensionless charge parameter $q$, appearing in \ref{2-2}, is inherited from the presence of higher dimensions and hence can assume both positive and negative values. This is in sharp contrast with the case of Maxwell field, where the corresponding contribution to the metric elements is through $Q^{2}$, which is strictly positive, irrespective of the sign of the electric charge $Q$. For positive values of $q$, \ref{2-2} appears similar to that of a \KN black hole with an event horizon and a Cauchy horizon, while the case with negative $q$ has no analogue in \gr\ and thus provides a true signature of the additional spatial dimensions \cite{Aliev:2005bi}. In particular, note that for negative $q$, it is possible to have a rotating black hole with its dimensionless angular momentum larger than unity, since the condition for turning the black hole extremal is given by, $(a/M)^{2}=1-q$. This provides yet another distinct test of the presence of extra dimensions, which we will comment upon in the subsequent discussions. This provides the basic framework within which we will be working in this paper, with the line element presented in \ref{2-2} playing the pivotal role. In what follows we will provide an analytical estimation of the epicyclic frequencies associated with the rotating braneworld black hole, which will be used subsequently to estimate the various QPOs observed in black hole spacetimes.

%%%%%%%%%%%%%%%%%%%%%%%%%%%%%%%%%%%%%%%%%%%%%%%%%%%%%%%%%%%%%%%%%%%%%%%%%%%%%%%%%%%%%%%%%%%%%%%%%%%
%%%%%%%%%%%%%%%%%%%%%%%%%%%%%%%%%%%%%%%%%%%%%%%%%%%%%%%%%%%%%%%%%%%%%%%%%%%%%%%%%%%%%%%%%%%%%%%%%%%
%%%%%%%%%%%%%%%%%%%%%%%%%%%%%%%%%%%%%%%%%%%%%%%%%%%%%%%%%%%%%%%%%%%%%%%%%%%%%%%%%%%%%%%%%%%%%%%%%%%
\section{Epicyclic frequencies of a rotating braneworld black hole}\label{S3}

In this section we will present the derivation of the epicyclic frequencies for the stationary, axisymmetric metric, depicting the spacetime around a rotating braneworld black hole. The epicyclic frequencies are most straightforward to obtain, if the spacetime is symmetric about the equatorial plane, which is true for the spacetime depicted by \ref{2-2}, since it can be expressed in the form,
%%%%%%%%%%%%%%%%%%%%%%%%%%%%%%%%%%%%%%%%%%%%%%%%%%%%%%%%%%%%%%%%%%%
\begin{align}
ds^2=g_{tt}dt^2 + 2g_{t\phi}dt d\phi + g_{\phi\phi}d\phi ^2 + g_{rr}dr^2 +g_{\theta\theta} d\theta^2~,
\label{3-1}
\end{align}
%%%%%%%%%%%%%%%%%%%%%%%%%%%%%%%%%%%%%%%%%%%%%%%%%%%%%%%%%%%%%%%%%%%
with $g_{\mu\nu}=g_{\mu\nu}(r,\theta)$ and $g_{\mu\nu}(r,\theta)=g_{\mu\nu}(r,-\theta)$, implying reflection symmetry of the metric. The epicyclic frequencies originate from the relaxation of the circular orbits under external perturbations. Since the accretion disc effectively constitutes of a large number of material particles moving in circular orbits of varied radius in the black hole spacetime. Owing to the existence of two Killing vector fields $(\partial/\partial t)^{\mu}$ and $(\partial/\partial \phi)^{\mu}$, it follows that we have two conserved quantities --- (a) the energy per unit mass $E=-u_{\mu}(\partial/\partial t)^{\mu}$ and (b) the angular momentum per unit mass $L=u_{\mu}(\partial/\partial \phi)^{\mu}$, respectively. In terms of these conserved quantities, the circular orbits ($r=r_{\rm c},\dot{r}=0$) of test particles in the equatorial plane ($\theta=\pi/2$) experience the following effective potential, 
%%%%%%%%%%%%%%%%%%%%%%%%%%%%%%%%%%%%%%%%%%%%%%%%%%%%%%%%%%%%%%%%%%%%%%%%%
\begin{align}
\mathcal{V}(r)=-\left(\frac{1+E^{2}U(r_{c},\frac{\pi}{2})}{g_{rr}}\right)~;
\qquad
U(r,\theta)=g^{tt}-2\left(\frac{L}{E}\right)g^{t\phi}+\left(\frac{L}{E}\right)^{2}g^{\phi\phi}~,
\label{3-5}
\end{align}
%%%%%%%%%%%%%%%%%%%%%%%%%%%%%%%%%%%%%%%%%%%%%%%%%%%%%%%%%%%%%%%%%%%%%%%%%
The specific energy $E$ and specific angular momentum $L$ of these circular orbits are obtained by solving the following two relations, namely $\mathcal{V}(r)=0=(d\mathcal{V}/dr)$. The resulting expressions are in terms of $\Omega \equiv (d\phi/dt)$, which is the angular velocity of the massive particle moving in the circular trajectory and satisfies the following algebraic equation,
%%%%%%%%%%%%%%%%%%%%%%%%%%%%%%%%%%%%%%%%%%%%%%%%%%%%%%%%%%
\begin{align}
\label{3-8}
\frac{\partial g_{tt}}{\partial r}+2\Omega \frac{\partial g_{t\phi}}{\partial r}+\Omega^{2}\frac{\partial g_{\phi\phi}}{\partial r}=0~.
\end{align}
%%%%%%%%%%%%%%%%%%%%%%%%%%%%%%%%%%%%%%%%%%%%%%%%%%%%%%%%%%
Being quadratic, the above algebraic equation can be immediately solved, thus expressing the angular velocity $\Omega$ in terms of metric coefficients, such that,
%%%%%%%%%%%%%%%%%%%%%%%%%%%%%%%%%%%%%%%%%%%%%%%%%%%%%%%%
\begin{align}\label{3-9}
\Omega=\frac{-\partial_{r}g_{t\phi}\pm \sqrt{\left(-\partial_{r}g_{t\phi}\right)^{2}-\left(\partial_{r}g_{\phi\phi}\right) \left(\partial_{r}g_{tt}\right)}}{\partial_{r}g_{\phi\phi}}\equiv 2\pi \nu_\phi~,
\end{align}
%%%%%%%%%%%%%%%%%%%%%%%%%%%%%%%%%%%%%%%%%%%%%%%%%%%%%%%%
where the positive (negative) sign in \ref{3-9} refers to the co-rotating (counter-rotating) orbits respectively. Finally the last equality in the above expression defines the quantity $\nu_{\phi}$. This provides the frequency in which the massive objects circle around the black hole spacetime in a circular orbit. 

As emphasized before, derivation of the epicyclic frequencies associated with circular orbits require one to perturb the circular orbits slightly in both the radial and the vertical directions, such that,
%%%%%%%%%%%%%%%%%%%%%%%%%%%%%%%%%%%%%%%%%%%%%%%%%%%%%%%%%%%%%%%%
\begin{align}
r(t)\simeq r_{\rm c}+\delta r_{0}~e^{i\omega_r t}~;
\qquad
\theta(t)\simeq \frac{\pi}{2}+\delta\theta_{0}~e^{i\omega_\theta t}~.
\label{3-10}
\end{align}
%%%%%%%%%%%%%%%%%%%%%%%%%%%%%%%%%%%%%%%%%%%%%%%%%%%%%%%%%%%%%%%%%
where $r_{\rm c}$ denotes the radius of the circular orbit which is being perturbed. Substitution of the time evolution of these radial and angular variables in the geodesic equation, yields the following estimations for the epicyclic frequencies, $\nu_{r}=(\omega_{r}/2\pi)$ and $\nu_{\theta}=(\omega_{\theta}/2\pi)$, as,
%%%%%%%%%%%%%%%%%%%%%%%%%%%%%%%%%%%%%%%%%%%%%%%%
\begin{align}
\label{3-13}
\nu_{r}^{2}=\frac{c^6}{G^2M^2}\Bigg[\frac{(g_{tt}+g_{t\phi}\Omega)^2}{2(2\pi)^2g_{rr}}\left(\frac{\partial^2{U}}{\partial r^2}\right)_{r_{\rm c},\frac{\pi}{2}}\Bigg]~,
\end{align}
%%%%%%%%%%%%%%%%%%%%%%%%%%%%%%%%%%%%%%%%%%%%%%%%
and
%%%%%%%%%%%%%%%%%%%%%%%%%%%%%%%%%%%%%%%%%%%%%%%%
\begin{align}
\label{3-14}
\nu_{\theta}^{2}=\frac{c^6}{G^2M^2}\Bigg[\frac{\big(g_{tt}+g_{t\phi}\Omega\big)^2}{2(2\pi)^2g_{\theta\theta}}\left(\frac{\partial^2{U}}{\partial \theta^2}\right)_{r_{\rm c},\frac{\pi}{2}}\Bigg] ~.
\end{align}
%%%%%%%%%%%%%%%%%%%%%%%%%%%%%%%%%%%%%%%%%%%%%%%%
The factors involving $(c^6/G^2M^2)$ have been incorporated in the expressions for radial and vertical epicyclic frequencies, so that they have the appropriate dimensions of frequency squared. Here $U(r,\theta)$ is the function appearing in the effective potential, whose expression is given in terms of the metric coefficients by \ref{3-5}. Note that the dependence of these epicyclic frequencies on the tidal charge $q$ is through the effective potential $U(r,\theta)$, but also through the combination $(g_{tt}+\Omega g_{t\phi})$, as well as $g_{\theta \theta}$ and $g_{rr}$. Since the explicit expressions of these quantities in terms of the tidal charge parameter is involved, we will not provide such explicit expressions in this work, rather shall provide numerical estimations of these frequencies and comparison with observed data.  

%%%%%%%%%%%%%%%%%%%%%%%%%%%%%%%%%%%%%%%%%%%%%%%%%%%%%%%%%%%%%%%%%%%%%%%%%%%%%%%%%%%%%%%%%%%%%%%%%%%
%%%%%%%%%%%%%%%%%%%%%%%%%%%%%%%%%%%%%%%%%%%%%%%%%%%%%%%%%%%%%%%%%%%%%%%%%%%%%%%%%%%%%%%%%%%%%%%%%%%
%%%%%%%%%%%%%%%%%%%%%%%%%%%%%%%%%%%%%%%%%%%%%%%%%%%%%%%%%%%%%%%%%%%%%%%%%%%%%%%%%%%%%%%%%%%%%%%%%%%
\section{Various Models for Quasi-Periodic Oscillation: Implications for braneworld black holes}\label{S4}

In this section we will spell out various theoretical models for QPOs, all of which effectively tries to explain the QPOs using various combinations of the epicyclic frequencies derived in the previous section. We will try to use these models along with the observed values from high frequency QPOs to understand if some constraint on the tidal charge parameter $q$ can be obtained. In particular, we will look for QPOs in stellar mass black holes with frequencies $\sim$ hundreds of Hz and also in supermassive BHs \cite{2008Natur.455..369G,Torok:2004xs,Aschenbach:2004kj} with frequencies $\sim $ mHz. The existence of mHz QPOs for supermassive BHs can be directly attributed to their larger masses compared to the stellar mass sources.

%%%%%%%%%%%%%%%%%%%%%%%%
%%%%%%%%%%%%%%%%%%%%%%%%
%%%%%%%%%%%%%%%%%%%%%%%%
%%%%%%%%%%%%%%%%%%%%%%%%
\begin{table}[h]
%\vskip0.2cm
\begin{center}
%\hspace*{-2.40cm}
\begin{tabular}{|c|c|c|c|c|}
\hline
$\rm Source $ & $\rm Mass$ & $\rm \nu_{up1} \pm \rm \Delta \nu_{up1}$ & $\rm \nu_{up2}\rm \pm \Delta \nu_{up2}$ & $\rm \nu_L \pm \Delta \nu_L$\\
%\hline
& $\rm (\rm in ~M_\odot)$ & $(\rm in~ Hz)$ & $\rm (\rm in ~Hz)$ & $\rm (\rm in~ Hz)$\\
\hline 
$\rm GRO ~J1655-40$ & $\rm 5.4\pm 0.3$ \cite{Beer:2001cg} & $\rm 441  \rm \pm 2 $ \cite{Motta:2013wga} & $\rm 298 \rm \pm 4 $ \cite{Motta:2013wga} & $\rm 17.3 \pm \rm 0.1 $ \cite{Motta:2013wga}\\ 
\hline
$\rm XTE ~J1550-564$ & $\rm 9.1\pm 0.61$ \cite{Orosz:2011ki} & $\rm 276 \rm \pm 3 $ & $\rm 184  \pm 5 $ & $ -$\\
\hline
%\vspace{0.2cm}
$\rm GRS ~1915+105$ & $\rm 12.4^{+2.0}_{-1.8}$ \cite{Reid:2014ywa} & $\rm 168  \pm 3 $ & $\rm 113  \pm 5 $ & $\rm - $\\
\hline
$\rm H ~1743+322$ & $\rm 8.0-14.07$ \cite{Pei:2016kka,Bhattacharjee:2019vyy,Petri:2008jc} & $\rm 242 \pm 3 $ & $\rm 166  \pm 5 $ & $\rm - $\\
\hline
$\rm Sgr~A^*$ & $\rm (3.5-4.9)$ & $\rm (1.445 \pm 0.16)$ & $\rm (0.886 \pm 0.04)$ & $ - $\\
%\hline
 & $\rm ~\times 10^6$ \cite{Ghez:2008ms,Gillessen:2008qv} & $\rm ~\times 10^{-3} $ \cite{Torok:2004xs,Stuchlik:2008fy} & $\rm ~\times 10^{-3} $ \cite{Torok:2004xs,Stuchlik:2008fy} & $ - $\\
 \hline
 $\rm RE J1034+396$ & $\rm (1-4) ~\times 10^6$ & $\rm (2.5-2.8) \rm ~\times 10^{-4}$ & $-$ & $ - $\\
%\hline
 &  \cite{2008Natur.455..369G,2012MNRAS.420.1825J,Czerny:2016ajj,Chaudhury:2018jzz} &  \cite{2008Natur.455..369G,Middleton:2008fe,Jin:2020hgq,Jin:2020meg} &   & \\
\hline
\end{tabular}
\caption{Black hole sources exhibiting high frequency QPOs (HFQPOs)}
\label{Table1}
\end{center}

\end{table}
%%%%%%%%%%%%%%%%%%%%%%%%
%%%%%%%%%%%%%%%%%%%%%%%%
%%%%%%%%%%%%%%%%%%%%%%%%
%%%%%%%%%%%%%%%%%%%%%%%%

In \ref{Table1} we have presented a few BH sources where QPOs have been discovered. As evident, all the HFQPOs for stellar mass black holes are in the range of hundreds of Hz, while for the supermassive BHs the QPOs are in the mHz range. Moreover, the QPOs are generally observed in commensurable pairs ($\nu_{\rm up1}$ and $\nu_{\rm up2}$) often with the ratio 3:2 (see \ref{Table1}). In this regard RE J1034+396 galaxy, the first AGN source with a significant detection of QPO in X-rays is an exception, since it exhibits only a single QPO in its power spectrum \cite{2008Natur.455..369G,Middleton:2008fe,Jin:2020hgq,Jin:2020meg}. 
Several theoretical models have been proposed to explain the commensurability of QPO frequencies (particularly the 3:2 ratio) \cite{Stella:1997tc,PhysRevLett8217,Stella_1999,Cadez:2008iv,Kostic:2009hp,Germana:2009ce,Kluzniak:2002bb,Abramowicz:2003xy,Rebusco:2004ba,Nowak:1996hg,Torok:2010rk,Torok:2011qy,Kotrlova:2020pqy,1980PASJ...32..377K,Perez:1996ti,Silbergleit:2000ck,Dexter:2013sxa} which mostly aim to explain the observed QPO frequencies in terms of the angular frequency $\nu_\phi$ and the epicyclic frequencies $\nu_r$ and $\nu_\theta$ introduced in \ref{S3}. Since these frequencies depend solely on the background spacetime and not on the complex physics of the accretion processes, QPOs can potentially be a more promising probe to test the nature of the background metric (or, to test the no-hair theorems) compared to other techniques, e.g., the iron line or the continuum-fitting methods.
The present work aims to discern the signatures of the tidal charge from the QPO signals in BHs using the available QPO models in the literature. Since these models primarily explain the 3:2 ratio BH HFQPOs, we will consider only those sources here where twin peak QPOs are observed (namely, the first five sources in \ref{Table1}). 

%%%%%%%%%%%%%%%%%%%%%%%%
%%%%%%%%%%%%%%%%%%%%%%%%
%%%%%%%%%%%%%%%%%%%%%%%%
%%%%%%%%%%%%%%%%%%%%%%%%
\begin{table}[h!]
\begin{center}
\begin{tabular}{|c|c|c|c|}
\hline
$\rm Model $ & $\rm \nu_{1} $ & $\rm \nu_{2}$ &  $\nu_3 $ \\
\hline 
$\rm Relativistic ~Precession ~Model ~(kinematic)$ \cite{Stella:1997tc,PhysRevLett8217,Stella_1999} & $\rm \nu_\phi$ & $\rm \nu_\phi-\nu_r $ & $\rm \nu_\phi-\nu_\theta$ \\ 
\hline
$\rm Tidal ~Disruption~Model~(kinematic)$ \cite{Cadez:2008iv,Kostic:2009hp,Germana:2009ce} & $\rm \nu_\phi + \nu_r$ & $\rm \nu_\theta $ & $-$ \\
\hline
$\rm Parametric ~Resonance~Model~(resonance)$ \cite{Kluzniak:2002bb,Abramowicz:2003xy,Rebusco:2004ba} & $\rm \nu_\theta$ & $\rm \nu_r$ & $-$\\ \hline
$\rm Forced ~Resonance~Model~1 ~(resonance)$ \cite{Kluzniak:2002bb} & $\rm \nu_\theta$ & $\rm \nu_\theta-\nu_r$ & $-$\\ \hline
$\rm Forced ~Resonance~Model~2 ~(resonance)$ \cite{Kluzniak:2002bb} & $\rm \nu_\theta+ \nu_r$ & $\rm \nu_\theta$ & $-$\\ \hline
$\rm Keplerian ~Resonance~Model~1 ~(resonance)$ \cite{Nowak:1996hg} & $\rm \nu_\phi$ & $\rm \nu_r$ & $-$\\ \hline
$\rm Keplerian ~Resonance~Model~2 ~(resonance)$ \cite{Nowak:1996hg} & $\rm \nu_\phi$ & $\rm 2\nu_r$ & $-$\\ \hline
$\rm Keplerian ~Resonance~Model~3 ~(resonance)$ \cite{Nowak:1996hg} & $\rm 3 \nu_r$ & $\rm \nu_\phi$ & $-$\\ \hline
$\rm Warped ~Disk~Oscillation~Model~ ~(resonance)$ & $\rm 2\nu_\phi-\nu_r$ & $\rm 2(\nu_\phi-\nu_r)$ & $-$\\ \hline
$\rm Non-axisymmetric ~Disk~Oscillation~Model~1 ~(resonance)$ & $\rm \nu_\theta$ & $\rm \nu_\phi-\nu_r$ & $-$\\ \hline
$\rm Non-axisymmetric ~Disk~Oscillation~Model~2 ~(resonance)$  \cite{Torok:2010rk,Torok:2011qy,Kotrlova:2020pqy} & $\rm 2\nu_\phi-\nu_\theta$ & $\rm \nu_\phi-\nu_r$ & $-$\\ \hline
\end{tabular}
\caption{Theoretical expressions for the HFQPOs and the LFQPO from various QPO models.}
\label{Table2}
\end{center}
\end{table}
%%%%%%%%%%%%%%%%%%%%%%%%
%%%%%%%%%%%%%%%%%%%%%%%%
%%%%%%%%%%%%%%%%%%%%%%%%
%%%%%%%%%%%%%%%%%%%%%%%%

Let us now briefly review some of the existing models explaining the QPOs by relating them to the angular frequency and the epicyclic frequencies presented in the previous section. This will be helpful in our subsequent investigation in assessing the prospect of these models in explaining the available observations associated with QPOs. For clarity, we denote the model-based HFQPOs (from different theoretical QPO models) by $\nu_1$ and $\nu_2$ respectively (analogous to $\nu_{\rm up1}$ and $\nu_{\rm up2}$ in \ref{Table1}). The theoretical expression for the low frequency QPO is represented by $\nu_3$ (analogous to $\nu_{\rm L}$ in \ref{Table1}). Following which we present a summary of the various QPO models and the analytical expressions for the HFQPOs, i.e., $\nu_1$ and $\nu_2$, in terms of angular and epicyclic frequencies introduced above, in \ref{Table2}. As emphasized, we note from \ref{Table2} that $\nu_1$ and $\nu_2$ in each of these models are some linear function of $\nu_\phi$, $\nu_r$ and $\nu_\theta$. Further, analytical estimation of these frequencies, presented in \ref{3-9}, \ref{3-13} and \ref{3-14}, respectively, suggest that these three frequencies depend on the radius from which these QPOs are being emitted, the mass of the black hole $M$ and the other black hole hairs, which in the present case of braneworld black hole corresponds to the spin parameter $a$ and the tidal charge parameter $q$. In this section, we intend to find out the allowed values of $q$ and $a$ from the available QPO observations of BHs, presented in \ref{Table1} in the context of each of the QPO models listed in \ref{Table2}. To accomplish the same, we adopt the following procedure:

%%%%%%%%%%%%%%%%%%%%%%%%
%%%%%%%%%%%%%%%%%%%%%%%%
%%%%%%%%%%%%%%%%%%%%%%%%
%%%%%%%%%%%%%%%%%%%%%%%%
\begin{enumerate}

\item We start by choosing a QPO model from \ref{Table2}, thereby fixing the expression for $\nu_1$ and $\nu_2$ (and also $\nu_3$ depending on the scope of the model) in terms of $\nu_{\phi}$ and epicyclic frequencies. These frequencies depend on the radius of the circular orbit $r_{\rm em}$ from which these QPOs have been generated, the black hole mass $M$, the tidal charge parameter $q$ and the rotation parameter $a$.

\item We next choose a black hole source from \ref{Table1}, fixing the range of mass allowed to us for computation of $\nu_1$ and $\nu_2$ (possibly, also $\nu_3$). 

\item Then we fix the tidal charge parameter $q$ to a certain value. Even though there are weaker constraints on the tidal charge parameter, say from the solar system tests, we will keep $q$ arbitrary for the time being, except for the bound $q\leq 1$. This originates from the demand of the existence of an event horizon.  

\item For the chosen value of the tidal charge $q$, the spin $a$ can be in the following range: $(-\sqrt{1-q})\leq a \leq \sqrt{1-q}$, such that the event horizon, located at $r_{\rm H}=1+\sqrt{1-a^2-q}$ is real. Note that for $q<0$, the BHs can assume spin values greater than unity, a distinctive signature of higher dimensional theory.

\item We allow the mass $M$ of the chosen BH source to vary in its observed range $(M_{0}-\Delta M) \leq M \leq (M_{0}+\Delta M)$ (where $M_{0}$ is the centroid value and $\Delta M$ is the error in the mass measurement given in \ref{Table1}). For each value of the mass $M$, we vary the emission radius $r_{\rm em}$ within the range: $r_{\rm ms} (a,q)\leq r_{\rm em} \leq r_{\rm ms}(a,q)+20 ~R_{\rm g}$ \cite{Goluchova:2019tgq,Kotrlova:2020pqy}. For each such combination of $M$ and $r_{\rm em}$ we calculate $\nu_1$, $\nu_2$ (and also $\nu_3$ depending on the chosen model) keeping the $q$ fixed and varying $a$ in the range allowed by the given $q$ (as discussed in Step 4).

\item For the given tidal charge $q$, we note the values of $a$ which give ($\nu_{\rm up1}-\Delta \nu_{\rm up1})\leq \nu_1 \leq (\nu_{\rm up1} + \Delta \nu_{\rm up1}$) and ($\nu_{\rm up2}-\Delta \nu_{\rm up2})\leq \nu_2 (\leq \nu_{\rm up2} + \Delta \nu_{\rm up2}$ )
(and also $(\nu_{\rm L}-\Delta \nu_{\rm L})\leq \nu_3 \leq (\nu_{\rm L} + \Delta \nu_{\rm L})$, if observed in the source and addressed by the model).

\item Steps 3-6 is repeated for the same source and the chosen QPO model for a different value of the tidal charge $q$. When this is done for a certain range of $q$, an area is traced in the $q-a$ plane for the chosen source. This area represents the values of $q$ and $a$ for the given source that can explain its observed QPO frequencies within the domain of the chosen model.

\item We then repeat the steps 2 to 7 for all the sources in \ref{Table1}, keeping the QPO model fixed.

\item Then we move to the next QPO model of \ref{Table2} and repeat the steps 1 to 8. This has generated \ref{Fig_01}, \ref{Fig_02}, \ref{Fig_03}, \ref{Fig_04}, \ref{Fig_05} and \ref{Fig_06} in the context of the various QPO models mentioned in \ref{Table2}, which we will discuss next.
\end{enumerate}
%%%%%%%%%%%%%%%%%%%%%%%%
%%%%%%%%%%%%%%%%%%%%%%%%
%%%%%%%%%%%%%%%%%%%%%%%%
%%%%%%%%%%%%%%%%%%%%%%%%

We must emphasize that our analysis tacitly assumes that the theoretical (model-based) QPO frequencies $\nu_1$ and $\nu_2$ (and $\nu_3$) are generated at the same circular orbit, with radius $r_{\rm em}$ \cite{2016A&A...586A.130S,Yagi:2016jml,Kotrlova:2020pqy}. This holds true for the kinematic and the resonant models, discussed later in this paper. Apart from these models there also exist certain diskoseismic models which assume that the oscillatory modes leading to the twin HFQPOs are emitted at different radii of the accretion disk \cite{1980PASJ...32..377K,Perez:1996ti,Silbergleit:2000ck}. However, results based on magnetohydrodynamic simulations \cite{Tsang:2008fz,Fu:2008iw,Fu:2010tf} reveal that such models cannot adequately reproduce the 3:2 ratio of HFQPOs and we shall not consider such models in this work. The QPO model proposed by Dexter \& Blaes \cite{Dexter:2013sxa,Blaes:2011cz} assumes that the two observed HFQPOs correspond to the vertical epicyclic oscillation and the lowest order acoustic breathing mode in a radiation dominated flow and can successfully reproduce the observed 3:2 ratio. This model is based on the semi-analytic thin accretion disc model proposed by Agol \& Krolik \cite{Agol:1999dn} which is a generalization of the standard thin disc model \cite{Novikov:1973kta,Shakura:1972te,Page:1974he}. All these models assume the background spacetime to be given by the Kerr metric. Extending these models for the present background is beyond the scope of this work and shall be reported elsewhere. In the next two subsections we shall discuss the various kinematic and resonant models mentioned in \ref{Table2} to explain the observed quasi-periodic oscillations. 

%%%%%%%%%%%%%%%%%%%%%%%%%%%%%%%%%%%%%%%%%%%%%%%%%%%%%%%%%%%%%%%%
%%%%%%%%%%%%%%%%%%%%%%%%%%%%%%%%%%%%%%%%%%%%%%%%%%%%%%%%%%%%%%%%
%%%%%%%%%%%%%%%%%%%%%%%%%%%%%%%%%%%%%%%%%%%%%%%%%%%%%%%%%%%%%%%%
%%%%%%%%%%%%%%%%%%%%%%%%%%%%%%%%%%%%%%%%%%%%%%%%%%%%%%%%%%%%%%%%
\subsection{Kinematic Models}\label{Sec4-1}

In these models local motion of plasma in the accretion disk, around the central compact object, is considered as the origin of the QPOs. Among the possible kinematic models the most prominent one is the Relativistic Precession Model (henceforth referred to as RPM)\cite{Stella:1997tc,PhysRevLett8217}, originally proposed to explain the twin HFQPOs in neutron star sources, subsequently has been extended for black hole sources as well \cite{Stella_1999}. In the RPM, the observed upper and lower HFQPOs, with frequencies $\nu_{\rm up1}$ and $\nu_{\rm up2}$, are related to the orbital angular frequency and the epicyclic frequencies, such that $\nu_1=\nu_\phi$ and $\nu_2=\nu_\phi-\nu_r$. Intriguingly, RPM also attempts to address the observed LFQPO in BHs, with frequency $\nu_{\rm L}$, to $\nu_{3}=\nu_{\phi}-\nu_{\theta}$. Here $\nu_{\theta}$ is the vertical epicyclic frequency and all the QPOs have been assumed to emerge from the same radial distance $r_{\rm em}$, which is the radius of nearly circular orbits associated with accreting matter. 

\begin{figure}[htp]
%\hspace{-1.8cm}
\subfloat[Relativistic Precession Model \label{Fig_1a}]{\includegraphics[scale=0.45]{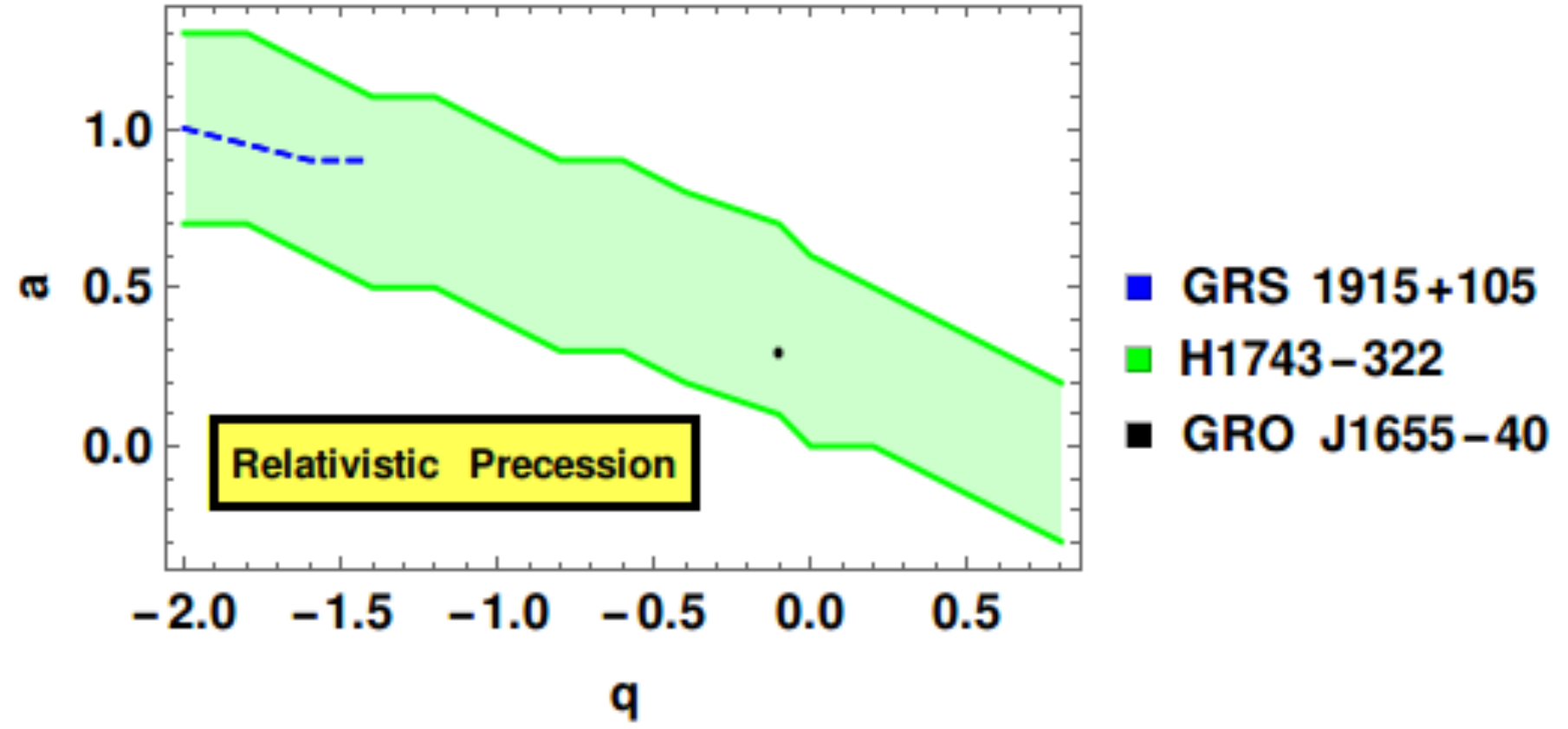}}
\subfloat[Tidal Disruption Model\label{Fig_1b}]{\includegraphics[scale=0.45]{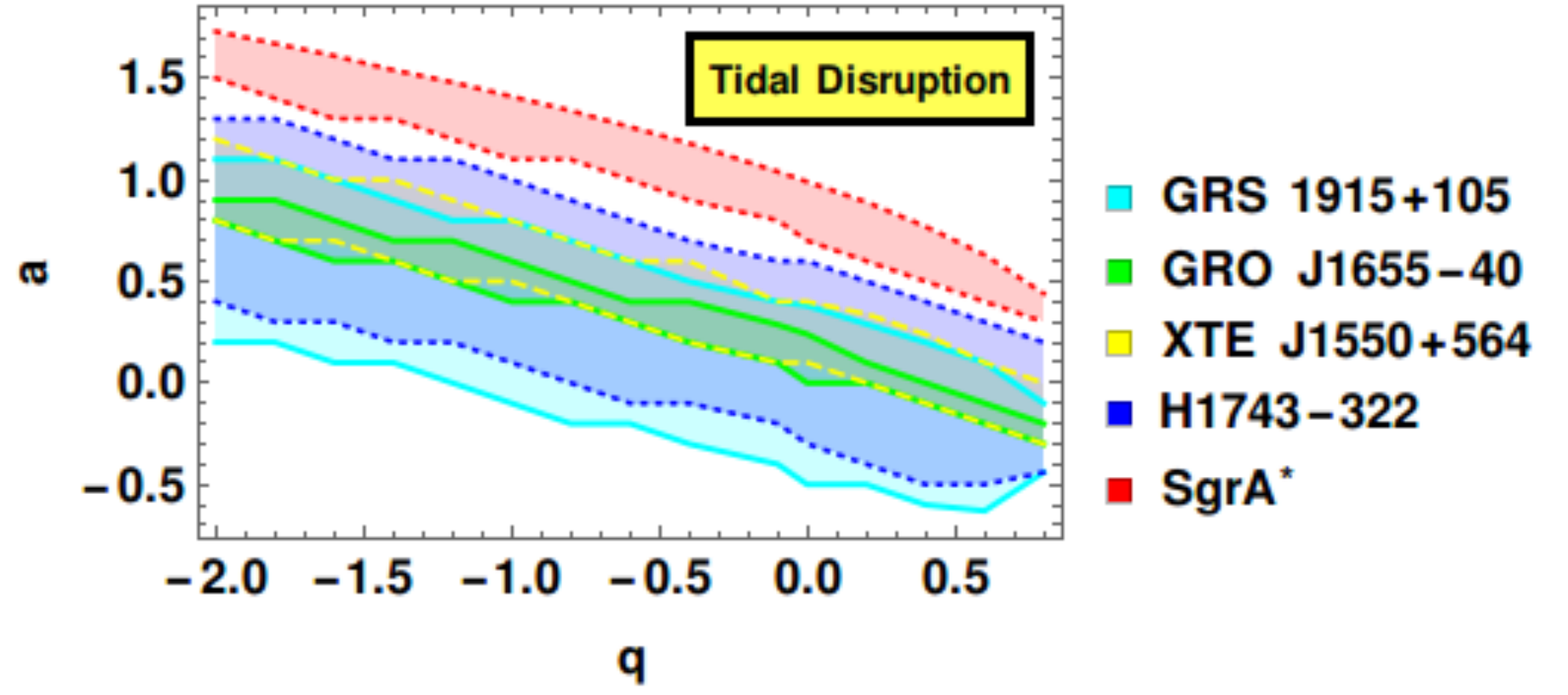}}
\caption{The above figure depicts the values of the tidal charge $q$ and the rotation parameter $a$ that can reproduce the observed QPO frequencies within error bars for different BH sources listed in \ref{Table1} assuming --- (a) the Relativistic Precession Model and (b) the Tidal Disruption Model. For more discussions see the main text.}
\label{Fig_01}
\end{figure}

As emphasized earlier, all these frequencies depend not only on the radius $r_{\rm em}$, at which these frequencies are excited, but also on the BH hairs. Therefore, if a BH source exhibits all the three QPO frequencies in its power spectrum, then these can be used to determine the hairs of the BH, which in the present context corresponds to $q$, $a$ and $M$, along with an estimation for the emitting radius $r_{\rm em}$. However, as emphasized earlier, in this work we do not attempt to constrain the mass of the BHs from the QPO observations, rather we will use the previously estimated mass of the BHs from other independent observations, e.g., optical/NIR photometry (see \ref{Table1}). Following the outline presented earlier in this section, the observationally allowed values of $q$ and $a$ for each of the BH sources, in the context of the RPM, has been presented in \ref{Fig_1a}. To summarize --- (a) Among the five sources presented in \ref{Table1}, the sources XTE J1550-564 and Sgr A* can not reproduce the observed twin HFQPOs, irrespective of the values of $q$ and $a$ within the RPM. (b) Further, as evident from \ref{Fig_1a}, the source GRO J1655-40 with $q\sim 0$ and $a\sim 0.3$ (represented by the black dot in \ref{Fig_1a}) can exactly reproduce all the three QPO frequencies \cite{Motta:2013wga}. (c) The observed HFQPOs (within the error bars) in GRS 1915+105 can be exactly explained if $-2\leq q \leq -1.5$ (blue dashed line in \ref{Fig_1a}), (d) while for the source H1743-322 the allowed values of $q$ and $a$ are denoted by the green shaded region in \ref{Fig_1a}. In order to get an estimation for the tidal charge parameter $q$ and the rotation parameter $a$, it is necessary to perform a joint-$\chi^2$ analysis with all the five sources present, which will be discussed in greater detail in \ref{S5}.

%%%%%%%%%%%%%%%%%%%%%%%%
%%%%%%%%%%%%%%%%%%%%%%%%
%%%%%%%%%%%%%%%%%%%%%%%%
%%%%%%%%%%%%%%%%%%%%%%%%
%%%%%%%%%%%%%%%%%%%%%%%%
%%%%%%%%%%%%%%%%%%%%%%%%
%%%%%%%%%%%%%%%%%%%%%%%%
%%%%%%%%%%%%%%%%%%%%%%%%

The Tidal Disruption Model \cite{Cadez:2008iv,Kostic:2009hp,Germana:2009ce} is another example of a kinematic model. According to this model, the plasma orbiting the BH may get tidally stretched by the central object, forming ring-like features along the orbit, which gives rise to the modulation in the observed flux and hence to the HFQPOs. Within the purview of this model, $\nu_1=\nu_\phi+\nu_r$ and $\nu_2=\nu_\phi$, which, as before, depend on the BH hairs $q$, $a$, and $M$ along with the emission radius $r_{\rm em}$. We compare $\nu_1$ and $\nu_2$ with $\nu_{\rm up1}\pm \Delta \nu_{\rm up1}$ and $\nu_{\rm up2}\pm \Delta \nu_{\rm up2}$, respectively, for each of the BH sources presented in \ref{Table1}. This enables us, following the flowchart presented before, to obtain the allowed values of $q$ and $a$ for each of these BH sources, which has been presented in \ref{Fig_1b}. We observe that the HFQPOs for all the five sources can be reproduced by all possible values of $q$. This indicates that the contribution to the joint-$\chi^2$ distribution from each of these sources for this model will be very close to zero. 

%%%%%%%%%%%%%%%%%%%%%%%%%%%%%%%%%%%%%%%%%%%%%%%%%%%%%%%%%%%%%%%%
%%%%%%%%%%%%%%%%%%%%%%%%%%%%%%%%%%%%%%%%%%%%%%%%%%%%%%%%%%%%%%%%
%%%%%%%%%%%%%%%%%%%%%%%%%%%%%%%%%%%%%%%%%%%%%%%%%%%%%%%%%%%%%%%%
%%%%%%%%%%%%%%%%%%%%%%%%%%%%%%%%%%%%%%%%%%%%%%%%%%%%%%%%%%%%%%%%
\subsection{Resonant Models}\label{Sec4-2}

The kinematic models, described above, cannot adequately explain the observed 3:2 ratio of the HFQPOs \cite{Torok:2011qy}. On the other hand, the resonant models \cite{2001A&A...374L..19A,Abramowicz:2003xy,2005A&A...436....1T} can explain the commensurability of the QPO frequencies more naturally. The fact that the observed twin-peak HFQPOs (with frequencies $\rm \nu_{up1}$ and $\rm \nu_{up2}$) in BH and NS sources occur in the ratio of 3:2, it clearly indicates that QPOs might result as a consequence of resonance between various oscillation modes in the accretion disk \cite{2001A&A...374L..19A,Kluzniak:2002bb,2001AcPPB..32.3605K}. In \ref{S3} we assumed that the matter in the accretion disk inspirals and then falls into the black hole, while maintaining nearly circular geodesics, along the equatorial plane. Such that slight perturbations $\delta r =r-r_{\rm c}$ and $\delta \theta=\theta-\pi/2$, from the circular radius $r_{\rm c}$ and from the equatorial plane, obey the equations of simple harmonic motion with frequencies $\omega_{r}$ and $\omega_{\theta}$, 
%%%%%%%%%%%%%%%%%%%%%%%%%%%%%%%%%%%%%%%%%%%%%%%%%%%%%%%%%%%%%%%%
\begin{align}
\delta \ddot{ r}+\omega_r^2 \delta r=0~; \qquad \delta \ddot{\theta}+\omega_\theta^2 \delta \theta=0~. 
\label{S4-1}
\end{align} 
%%%%%%%%%%%%%%%%%%%%%%%%%%%%%%%%%%%%%%%%%%%%%%%%%%%%%%%%%%%%%%%%
Here $\omega_{r}=2\pi \nu_r$ and $\omega_{\theta}=2\pi \nu_\theta$, with $\nu_{r}$ and $\nu_{\theta}$ are the radial and vertical epicyclic frequencies, as  discussed in \ref{S3} and the dot denotes derivative with respect to the coordinate time $t$. It is to be noted that, \ref{S4-1} represents two uncoupled harmonic oscillators, without any forcing terms, and applicable to thin disks as well as to more general accretion flow models, e.g. accretion tori \cite{2001A&A...374L..19A,Kluzniak:2002bb}. A more realistic model should however consider non-linear effects due to the pressure and dissipation in the accreting fluid, which requires inclusion of forcing terms in \ref{S4-1}, i.e.,
%%%%%%%%%%%%%%%%%%%%%%%%%%%%%%%%%%%%%%%%%%%%%%%%%%%%%%%%%%%%%%%%
\begin{align}
\delta \ddot{r}+\omega_r^2 \delta r=\omega_r^{2}F_{r}(\delta r,\delta \theta, \delta\dot{r}, \delta\dot{\theta})~; 
\qquad 
\delta \ddot{\theta}+\omega_\theta^2 \delta \theta=\omega_{\theta}^{2} F_{\theta}(\delta r,\delta \theta, \delta\dot{r}, \delta\dot{\theta})~.
\label{S4-2}
\end{align}
%%%%%%%%%%%%%%%%%%%%%%%%%%%%%%%%%%%%%%%%%%%%%%%%%%%%%%%%%%%%%%%%
In the above expressions, $F_{r}$ and $F_{\theta}$ are the forcing terms, which are in general some non-linear functions of their arguments. The forms of $F_r$ and $F_\theta$ depend on the accretion flow model, which aims to explain the observed QPOs. Since dissipation and pressure effects in the accretion physics that give rise to the forcing terms are not well understood, determining the correct analytical form of $F_{r}$ and $F_{\theta}$ is difficult. However, depending on the physical situation, one might be able to provide an analytical form for the forces $F_{r}$ and $F_{\theta}$, appearing in \ref{S4-2} \cite{Abramowicz:2003xy, Horak:2004hm}. We describe below various possible choices for the force terms, depending on the physics of the situation.  

%%%%%%%%%%%%%%%%%%%%%%%%
%%%%%%%%%%%%%%%%%%%%%%%%
%%%%%%%%%%%%%%%%%%%%%%%%
%%%%%%%%%%%%%%%%%%%%%%%%
\begin{itemize}

\item {{\bf Parametric Resonance Model:}} One of the simplest way to choose the forcing terms is to consider the situation when the vertical epicyclic mode is excited by the radial epicyclic mode, due to the fact that random fluctuations in thin disks are expected to have $\delta r \gg \delta \theta$ \cite{Kluzniak:2002bb,2001A&A...374L..19A,Abramowicz:2003xy,2005A&A...436....1T,Rebusco:2004ba}. Under these assumptions \ref{S4-2} take the form,
%%%%%%%%%%%%%%%%%%%%%%%%%%%%%%%%%%%%%%%%%%%%%%%%%%%%%%%%%%%%%%%%
\begin{align}
\delta \ddot{ r}+\omega_r^2 \delta r=0 ~~~~~~\delta \ddot{\theta}+\omega_\theta^2 \delta \theta= -\omega_\theta^2 \delta r \delta \theta
\label{S4-3}
\end{align}
%%%%%%%%%%%%%%%%%%%%%%%%%%%%%%%%%%%%%%%%%%%%%%%%%%%%%%%%%%%%%%%%
We note from \ref{S4-3} that $\delta r=B \cos (\omega_r t)$, such that the differential equation for $\delta\theta$ becomes,
%%%%%%%%%%%%%%%%%%%%%%%%%%%%%%%%%%%%%%%%%%%%%%%%%%%%%%%%%%%%%%%%
\begin{align}
\delta \ddot{\theta}+\omega_\theta^2 (1+B\delta r)\delta \theta =0 
\label{S4-4}
\end{align}
%%%%%%%%%%%%%%%%%%%%%%%%%%%%%%%%%%%%%%%%%%%%%%%%%%%%%%%%%%%%%%%%
where, $B$ is a constant. The above equation for $\delta \theta$, as presented in \ref{S4-4}, is akin to the Matthieu equation \cite{1969mech.book.....L}, which describes parametric resonance and is excited when \cite{Rebusco:2004ba,1969mech.book.....L}
%%%%%%%%%%%%%%%%%%%%%%%%%%%%%%%%%%%%%%%%%%%%%%%%%%%%%%%%%%%%%%%%
\begin{align}
\frac{\nu_r}{\nu_\theta}=\frac{2}{n}~, 
\qquad
\rm where ~ n \in ~positive ~integers~.
\label{S4-5}
\end{align} 
%%%%%%%%%%%%%%%%%%%%%%%%%%%%%%%%%%%%%%%%%%%%%%%%%%%%%%%%%%%%%%%%

%%%%%%%%%%%%%%%%%%%%%%%%
%%%%%%%%%%%%%%%%%%%%%%%%
%%%%%%%%%%%%%%%%%%%%%%%%
%%%%%%%%%%%%%%%%%%%%%%%%
\begin{figure}[h!]
\centering
\includegraphics[scale=0.65]{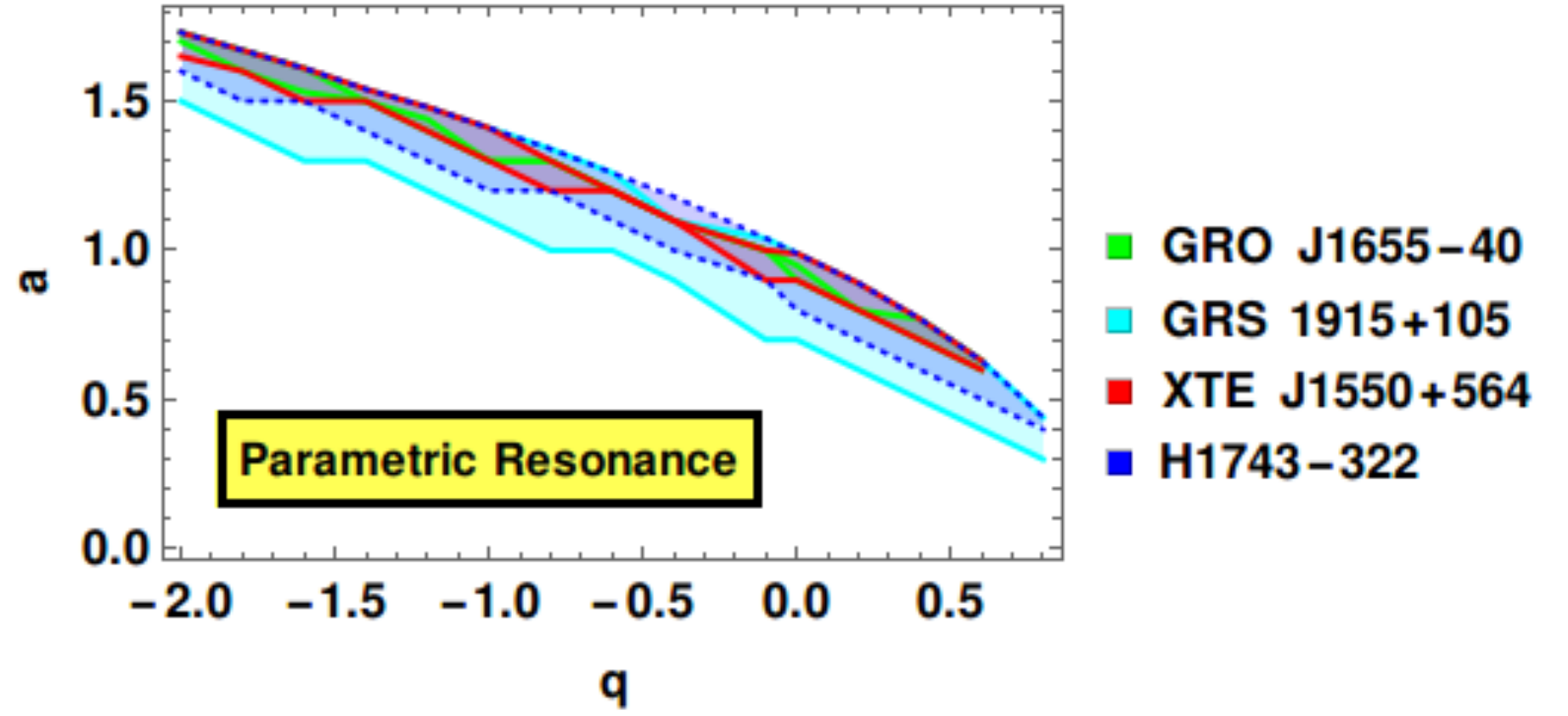}
\caption{The above figure depicts the values of $q$ and $a$ that can reproduce the observed QPO frequencies within the error bars for different BH sources (see \ref{Table1}) assuming the Parametric Resonance Model. See text for more discussions.}
\label{Fig_02}
\end{figure}
%%%%%%%%%%%%%%%%%%%%%%%%
%%%%%%%%%%%%%%%%%%%%%%%%
%%%%%%%%%%%%%%%%%%%%%%%%
%%%%%%%%%%%%%%%%%%%%%%%%

The resonance is strongest for the smallest values of $n$. Since for braneworld black holes $\nu_\theta > \nu_r$ \cite{Stuchlik:2008fy}, the minimum value of $n$ can be $3$, which naturally explains the observed $3:2$ ratio of the HFQPOs, such that $\nu_{1}=\nu_\theta$ and $\nu_{2}=\nu_r$. The fact that a coupling between the radial and vertical epicyclic frequencies can indeed give rise to the observed twin-peak HFQPOs in the ratio of $3:2$ was further confirmed in numerical simulations \cite{Abramowicz:2003xy} followed by analytical solutions \cite{Rebusco:2004ba,Horak:2004hm}.

In \ref{Fig_02} we have plotted the allowed values of the tidal charge $q$ and the rotation parameter $a$ for each of the black hole sources listed in \ref{Table1}. From the figure we note that the observed QPOs in SgrA* cannot be reproduced for any value of $q$, when Parametric Resonance model is considered, hence this source does not enclose any area in the $q-a$ plane in \ref{Fig_02}. On the other hand, the sources GRO J1655-40 and XTE J1550-564, for $q\geq 0.6$, cannot explain the observed HFQPOs, while all values of $q$ can reproduce the observed HFQPOs in GRS 1915+105 and H 1743-322. 

\item {\bf{ Forced Resonance Models:}} Parametric resonance seems to be quite natural in thin disks or nearly Keplerian accretion disks \cite{2001A&A...374L..19A,Kluzniak:2002bb,2001AcPPB..32.3605K}. In a more realistic accretion flow however, one may expect non-linear couplings between $\delta r$ and $\delta\theta$ in addition to parametric resonance. Such non-linear terms in $\delta r$ and $\delta\theta$ can be attributed to pressure, viscous or magnetic stresses present in the accretion flow such that $F_r$ and $F_\theta$ in \ref{S4-2} are non-zero \cite{2005A&A...436....1T}. Since the physics of the accretion flow is not very well understood, these forcing terms are often described by some mathematical ansatz. Lee et al. \cite{2004ApJ...603L..93L} showed in their numerical simulations that a resonant forcing of vertical oscillations by radial oscillations can be excited through a pressure coupling. Such a finding corroborates the earlier results of Abramowicz \& Kluzniak \cite{2001A&A...374L..19A} which models the HFQPOs in terms of a forced non-linear oscillator, such that, 
%%%%%%%%%%%%%%%%%%%%%%%%%%%%%%%%%%%%%%%%%%%%%%%%%%%%%%%%%%%%%%%%
\begin{align}
\delta \ddot{\theta}+\omega_\theta^2 \delta \theta =-\omega_\theta^2\delta r \delta \theta + \mathcal{F_\theta}(\delta\theta) 
\label{S4-7}
\end{align}
%%%%%%%%%%%%%%%%%%%%%%%%%%%%%%%%%%%%%%%%%%%%%%%%%%%%%%%%%%%%%%%%
where $\delta r=A cos(\omega_r t)$ and $\mathcal{F_\theta}$ denotes the non-linear terms in $\delta\theta$. \ref{S4-7} again allows solutions of the form:
%%%%%%%%%%%%%%%%%%%%%%%%%%%%%%%%%%%%%%%%%%%%%%%%%%%%%%%%%%%%%%%%
\begin{align}
\frac{\nu_\theta}{\nu_r}=\frac{m}{n} \rm~~~~~where ~m ~and~ n ~are~ natural~ numbers
\label{S4-8}
\end{align}
%%%%%%%%%%%%%%%%%%%%%%%%%%%%%%%%%%%%%%%%%%%%%%%%%%%%%%%%%%%%%%%%
While $m/n=3/2$ correspond to parametric resonance, the presence of non-linear terms allow the occurence of resonance between combinations of frequencies, e.g. $\nu_\theta-\nu_r$, $\nu_\theta+\nu_r$. Two such examples are $3:1$ and $2:1$ forced resonances where $m:n=3:1$ and $m:n=2:1$ respectively. 

%%%%%%%%%%%%%%%%%%%%%%%%
%%%%%%%%%%%%%%%%%%%%%%%%
%%%%%%%%%%%%%%%%%%%%%%%%
%%%%%%%%%%%%%%%%%%%%%%%%
\begin{figure}[h!]
%\hspace{-1.8cm}
\subfloat[3:1 Forced Resonance Model \label{Fig_3a}]{\includegraphics[scale=0.45]{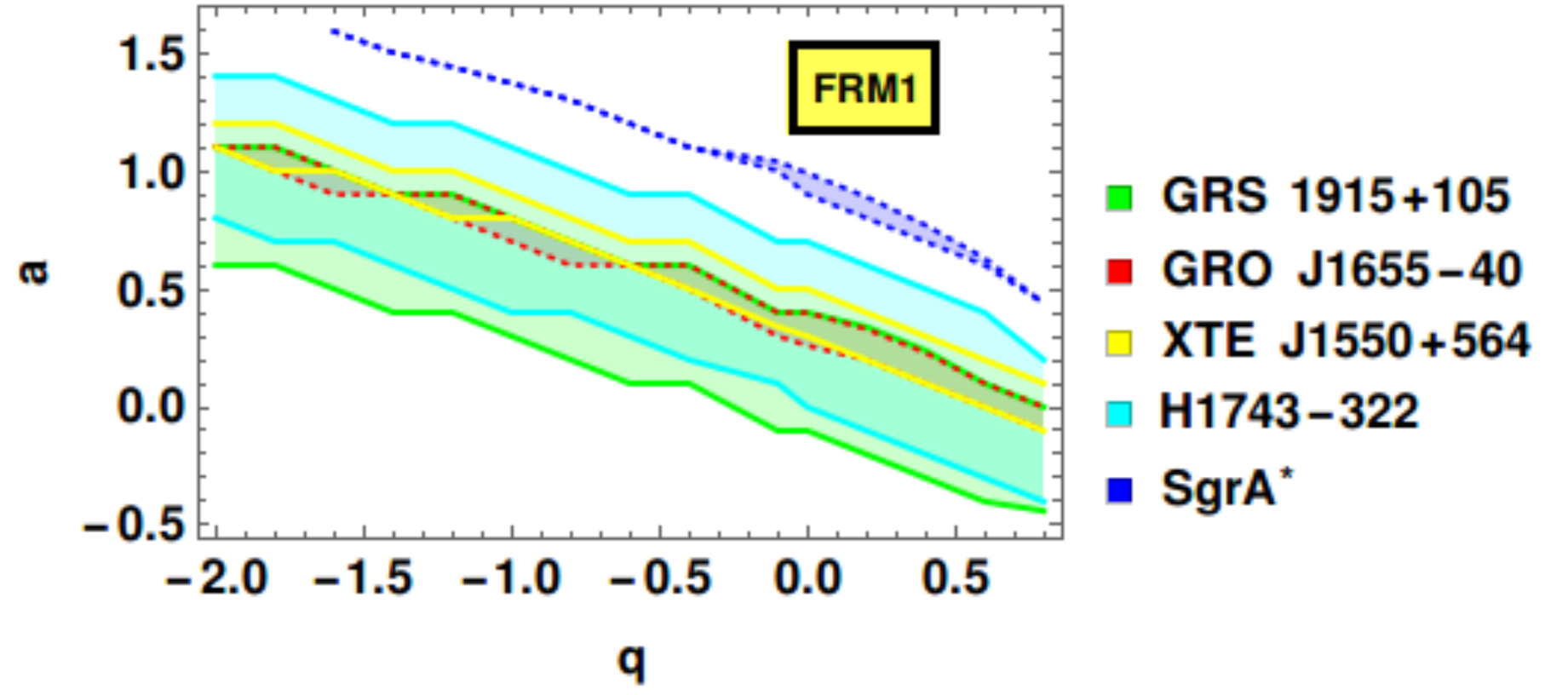}}
\subfloat[2:1 Forced Resonance Model\label{Fig_3b}]{\includegraphics[scale=0.45]{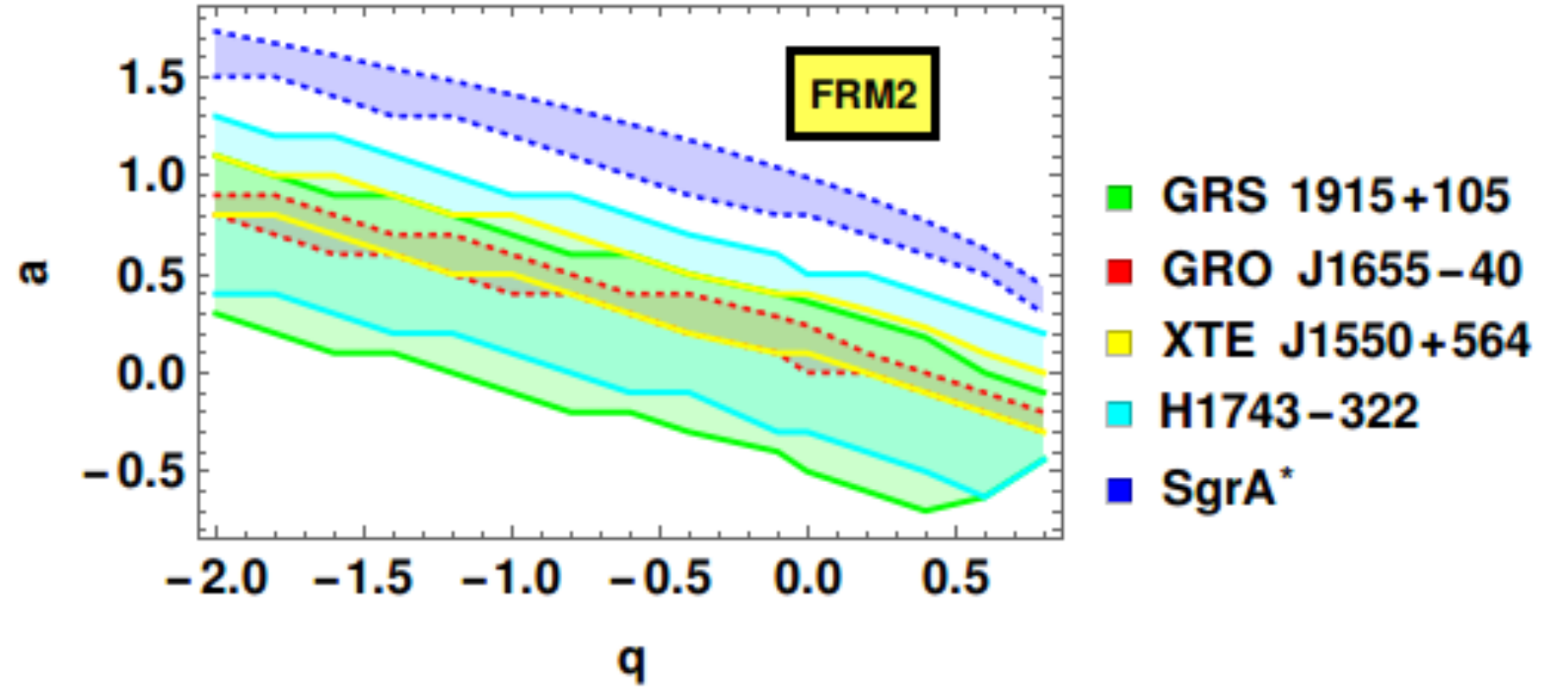}}
\caption{The above figure illustrates the values of $q$ and $a$ that can reproduce the observed QPO frequencies within the error bars for different BH sources (see \ref{Table1}) assuming the (a) 3:1 Forced Resonance Model (FRM1) and (b) 2:1 Forced Resonance Model (FRM2). For more discussions see main text.}
\label{Fig_03}
\end{figure}
%%%%%%%%%%%%%%%%%%%%%%%%
%%%%%%%%%%%%%%%%%%%%%%%%
%%%%%%%%%%%%%%%%%%%%%%%%
%%%%%%%%%%%%%%%%%%%%%%%%

In case of $3:1$ forced resonances (denoted by Forced Resonance Model 1 or FRM1) the upper HFQPO is given by $\nu_{1}=\nu_\theta$ while the lower HFQPO is given by $\nu_{2}=\nu_-= \nu_\theta-\nu_r$, i.e., 
%%%%%%%%%%%%%%%%%%%%%%%%%%%%%%%%%%%%%%%%%%%%%%%%%%%%%%%%%%%%%%%%
\begin{align}
\frac{\nu_\theta}{\nu_r}=\frac{m}{n}=3~~~\rm such ~ that~\nu_{1}=\nu_\theta=3\nu_r~~~while~\nu_{2}=\nu_-=\nu_\theta-\nu_r=2\nu_r
\label{S4-10}
\end{align}
%%%%%%%%%%%%%%%%%%%%%%%%%%%%%%%%%%%%%%%%%%%%%%%%%%%%%%%%%%%%%%%%
For $2:1$ forced resonances (denoted by Forced Resonance Model 2 or FRM2) $\nu_{1}=\nu_+= \nu_\theta+\nu_r$ while $\nu_{2}=\nu_\theta$, i.e, 
%%%%%%%%%%%%%%%%%%%%%%%%%%%%%%%%%%%%%%%%%%%%%%%%%%%%%%%%%%%%%%%%
\begin{align}
\frac{\nu_\theta}{\nu_r}=\frac{m}{n}=2~~~\rm such ~ that~\nu_{1}=\nu_+= \nu_\theta+\nu_r=3\nu_r~~~while~\nu_{2}=\nu_\theta=2\nu_r
\label{S4-9}
\end{align}
%%%%%%%%%%%%%%%%%%%%%%%%%%%%%%%%%%%%%%%%%%%%%%%%%%%%%%%%%%%%%%%%

Here \ref{Fig_3a} and \ref{Fig_3b} depict the observationally allowed values of $q$ and $a$ assuming FRM1 and FRM2 respectively, for each of the five BH sources mentioned in \ref{Table1}. For FRM2 (\ref{Fig_3b}) the observed twin peak QPOs in all the five BH sources can be addressed by all values of $q$. For FRM1 all values of $q$ can explain the HFQPOs in the microquasars while for SgrA* if $q<-1.6$ the observed QPO frequencies cannot be reproduced. From \ref{Fig_03} it is clear that the joint-$\chi^2$ cannot assume very high positive values in these models for reasons explained in \ref{Sec4-1}.

\item {{\bf Keplerian Resonance Models:}} For Parametric and Forced Resonance models the radial and vertical epicyclic frequencies are considered to be coupled. One can also consider non-linear resonances between the radial epicyclic modes and the orbital angular motion, which are characteristic to Keplerian Resonance models 
\cite{2005A&A...436....1T,2001PASJ...53L..37K,2001A&A...374L..19A, Nowak:1996hg}. Such a resonance may be physically realized in the inner region of relativistic thin accretion disks as a consequence of trapping of non-axisymmetric g-mode oscillations induced by a corotation resonance \cite{2001PASJ...53L..37K}. However, it was eventually realized that corotation resonance dampens the g-mode oscillations instead of exciting them and hence models invoking Keplerian resonance may not be very promising in explaing the HFQPOs in microquasars \cite{Li:2002yi,2003PASJ...55..257K}.  
  
%%%%%%%%%%%%%%%%%%%%%%%%
%%%%%%%%%%%%%%%%%%%%%%%%
%%%%%%%%%%%%%%%%%%%%%%%%
%%%%%%%%%%%%%%%%%%%%%%%%
\begin{figure}[h!]
%\hspace{-2cm}
%\centering
\subfloat[\label{Fig_4a}]{\includegraphics[scale=0.45]{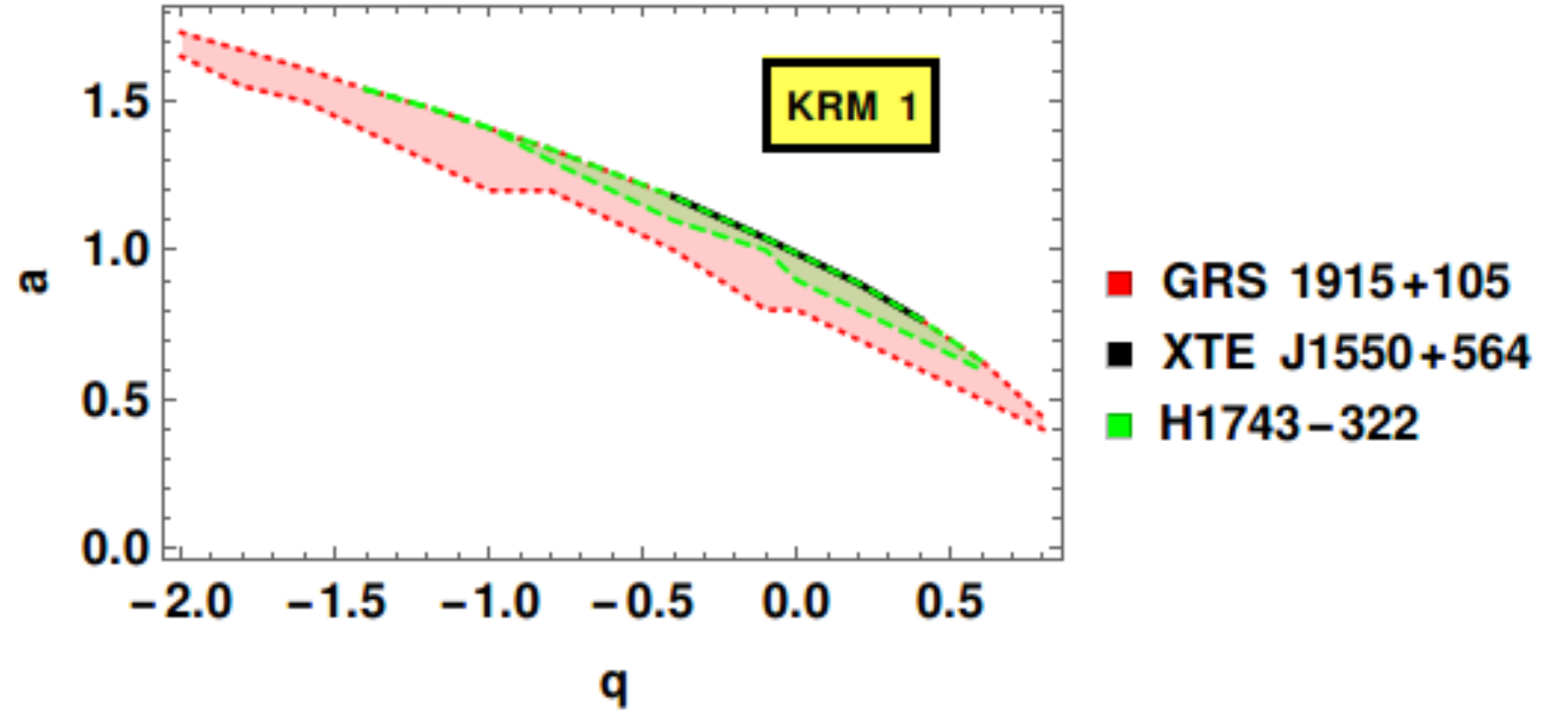}}~~
\subfloat[\label{Fig_4b}]{\includegraphics[scale=0.45]{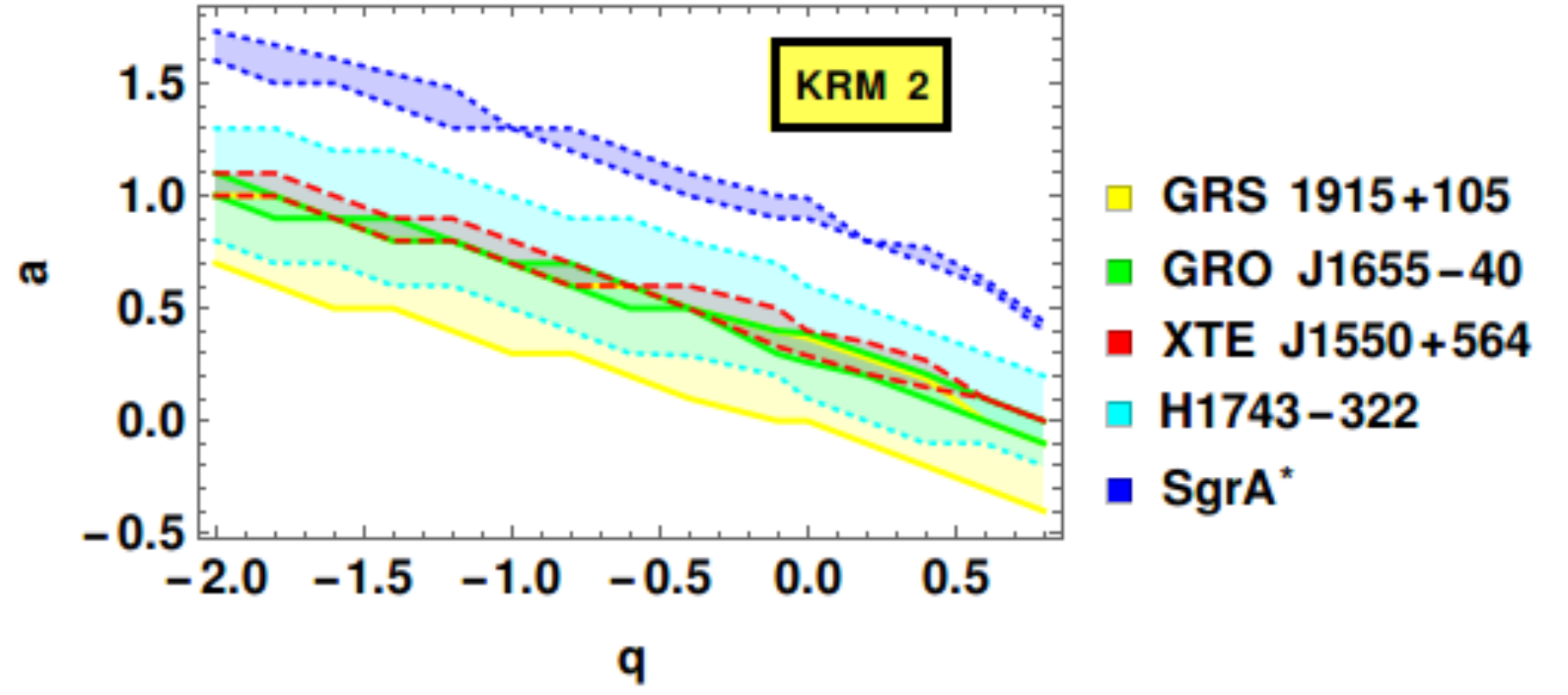}}\\
\centering
{\hspace{5cm}\subfloat[\label{Fig_4c}]{\includegraphics[scale=0.45]{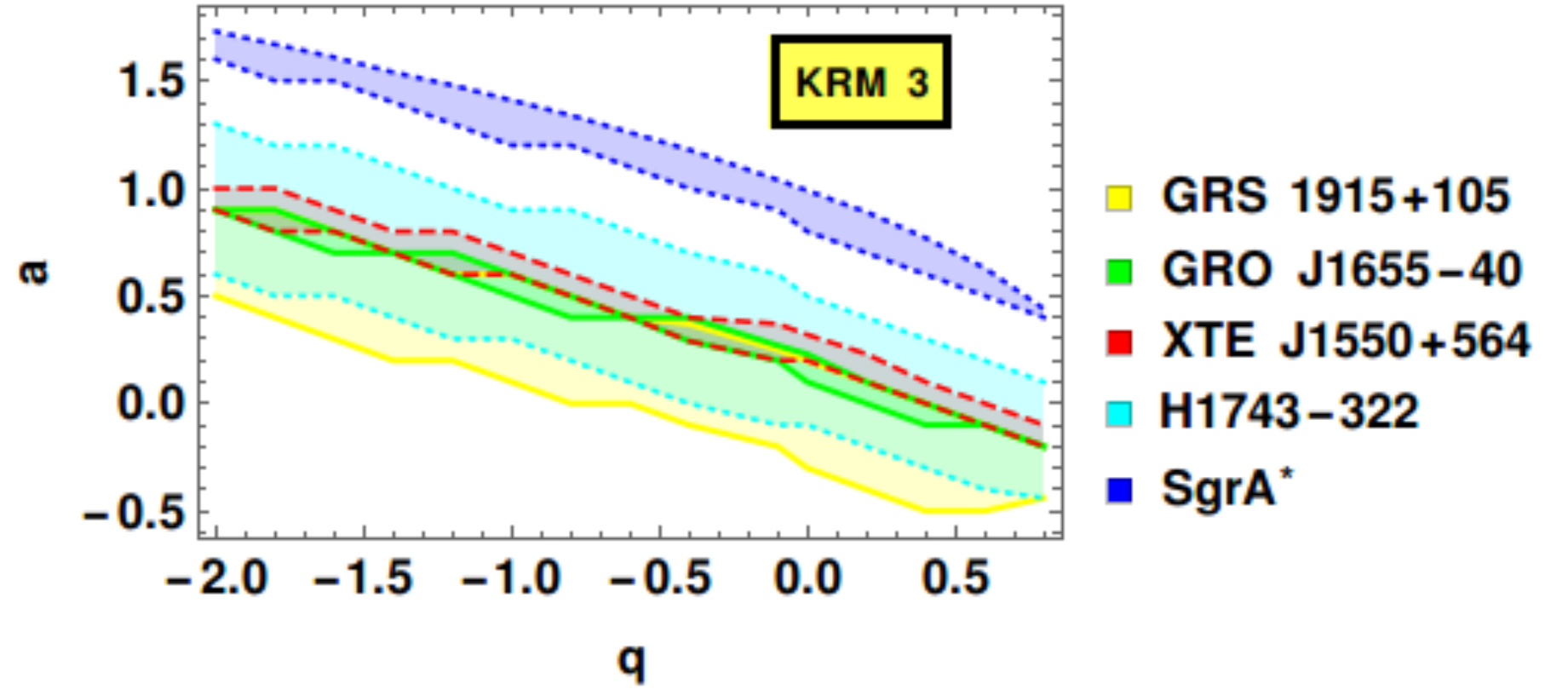}}}~~
\caption{The above figure illustrates the values of $q$ and $a$ that can explain the observed QPO frequencies within the error bars for different BH sources in \ref{Table1} assuming the (a) Keplerian Resonance Model 1 (KRM1), (b) Keplerian Resonance Model 2 (KRM2) and (c) Keplerian Resonance Model 3 (KRM3). For more discussions see text.}
\label{Fig_04}
\end{figure}
%%%%%%%%%%%%%%%%%%%%%%%%
%%%%%%%%%%%%%%%%%%%%%%%%
%%%%%%%%%%%%%%%%%%%%%%%%
%%%%%%%%%%%%%%%%%%%%%%%%

Another physical scenario where Keplerian resonance may occur corresponds to the situation when a pair of spatially separated coherent vortices with opposite vorticities oscillating with radial epicyclic frequencies couple with the  spatially varying orbital angular frequencies \cite{2005A&A...436....1T,2010tbha.book.....A}.
In order to explain the observed HFQPOs Keplerian resonance may occur between (i) $\nu_1=\nu_\phi$ and $\nu_2=\nu_r$ (which we denote as Keplerian Resonance Model 1 or KRM1), (ii) $\nu_1=\nu_\phi$ and $\nu_2=2\nu_r$ (which we denote as Keplerian Resonance Model 2 or KRM2) and (iii) $\nu_1=3\nu_r$  and $\nu_2=\nu_\phi$ (which we denote as Keplerian Resonance Model 3 or KRM3). 

In \ref{Fig_04} we present the allowed values of $q$ and $a$ for each of the black hole sources mentioned in \ref{Table1} in the context of Keplerian Resonance Model 1 (\ref{Fig_4a}), Keplerian Resonance Model 2 (\ref{Fig_4b}) and Keplerian Resonance Model 3 (\ref{Fig_4c}). We note from \ref{Fig_4b} and \ref{Fig_4c} that if KRM2 and KRM3 are considered then all values of $q$ can explain the observed HFQPOs in the five black holes. The joint-$\chi^2$ calculated from these models are expected to be small for all values of $q$ without much variation. Therefore KRM2 and KRM3 cannot be used to constrain the observationally preferred values of the tidal charge. Using Keplerian Resonance Model 1 (KRM 1) however, one can expect to see substantial variations in the joint-$\chi^2$ as $q$ is changed. This is mainly attributed to the fact that the observed HFQPOs of SgrA* and GRO J1655-40 cannot be reproduced by any allowed value of $q$ when KRM 1 is considered (\ref{Fig_4a}). Such a model can therefore establish stringent constraints on the tidal charge which we will explore in the next section. 

\item {{\bf Warped Disk Oscillation Model:}} This model considers non-linear resonances between various disk oscillation modes and the relativistic disk deformed by a warp \cite{2001PASJ...53....1K,2004PASJ...56..559K,2004PASJ...56..905K,2005PASJ...57..699K,2008PASJ...60..111K}. There are three types of resonances, namely, the horizontal resonances which can excite the p-mode and the g-mode oscillations and the vertical resonances which can induce only the g-mode oscillations \cite{2004PASJ...56..559K}. Such resonances occur due to the fact that in a relativistic disk the radial epicyclic frequency does not change monotonically with the radial distance $r$ \cite{2004PASJ...56..559K}. The drawback of this model is that it assumes a somewhat unusual disc geometry \cite{Torok:2011qy,Yagi:2016jml}. 

\ref{Fig_05} shows that all values of $q$ can reproduce the observed HFQPOs in all the five black hole sources when Warped Disk Oscillation Model (WDOM) is considered. Such a model therefore, cannot impose strong constraints on the observationally favored values of the tidal charge $q$.  

%%%%%%%%%%%%%%%%%%%%%%%%
%%%%%%%%%%%%%%%%%%%%%%%%
%%%%%%%%%%%%%%%%%%%%%%%%
%%%%%%%%%%%%%%%%%%%%%%%%
\begin{figure}[t!]
\centering
\includegraphics[scale=0.65]{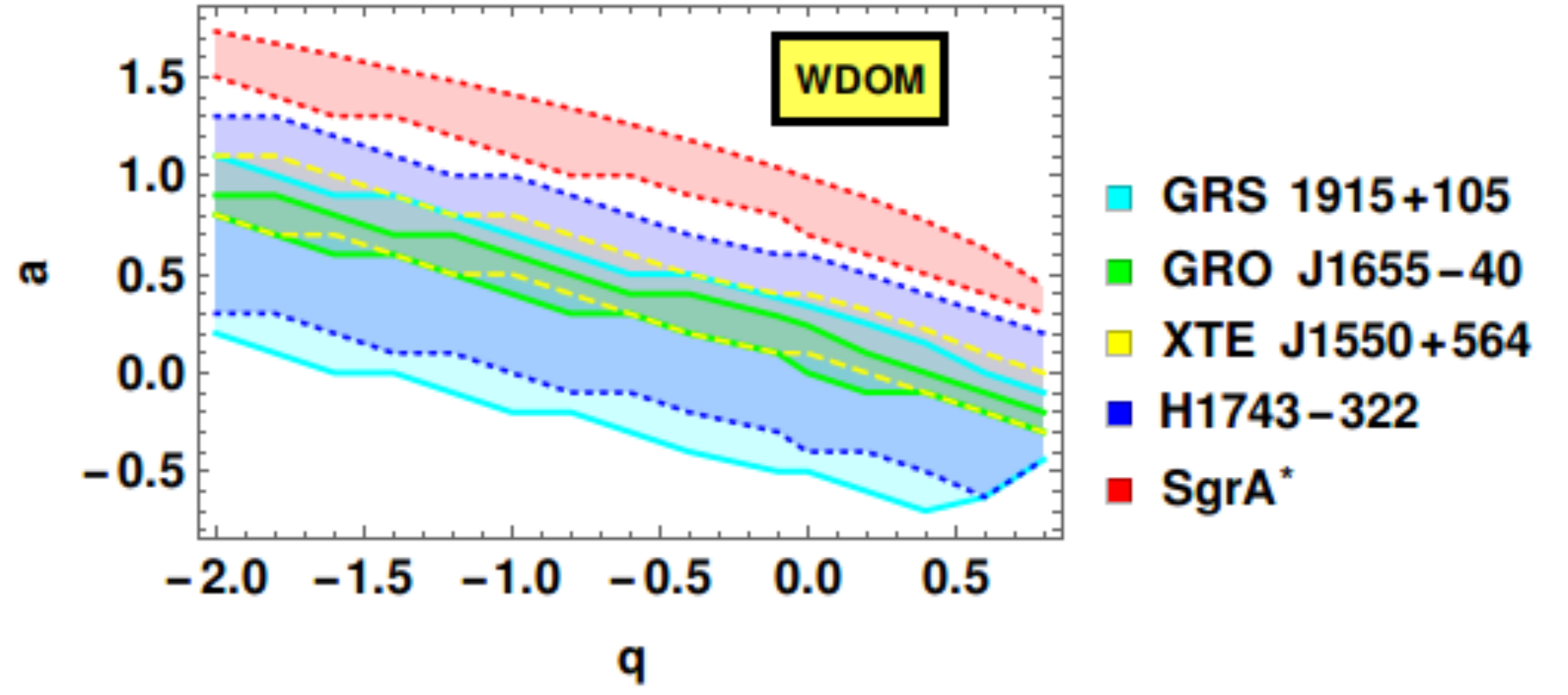}
\caption{The above figure depicts the values of $q$ and $a$ that can address the observed QPO frequencies within the error bars for different BH sources (see \ref{Table1}) assuming the Warped Disk Oscillation Model. See text for more discussions.}
\label{Fig_05}
\end{figure}
%%%%%%%%%%%%%%%%%%%%%%%%
%%%%%%%%%%%%%%%%%%%%%%%%
%%%%%%%%%%%%%%%%%%%%%%%%
%%%%%%%%%%%%%%%%%%%%%%%%

\item {{\bf Non-axisymmetric Disk-Oscillation Model:}} These models assume various combinations of non-axisymmetric disc oscillation modes as the origin of the HFQPOs \cite{2004ApJ...617L..45B,2005AN....326..849B,2005ragt.meet...39B,Torok:2011qy,Kotrlova:2020pqy}. 
They are essentially variants of the Relativistic Precession model since the oscillation modes they consider are associated with frequencies which nearly coincide with the predictions of RPM for slowly rotating black holes \cite{Torok:2011qy}. In these models the accretion flow is modeled in terms of a slightly non-slender pressure-supported perfect fluid torus such that the non-geodesic effects in the accretion flow originating from the pressure forces can be incorporated \cite{Torok:2015tpu,Sramkova:2015bha,Kotrlova:2020pqy}. 

The first variant of the non-axisymmetric disk oscillation model (referred here as NADO1) 
assumes resonance between the $m=-1$ non-axisymmetric radial epicyclic frequency ($\nu_2=\nu_{per}=\nu_\phi-\nu_r$) with the vertical epicyclic frequency ($\nu_1=\nu_\theta$) (here $m$ is the azimuthal wave number of the non-axisymmetric perturbation). This model, also known as the Vertical Precession Resonance Model \cite{2004ApJ...617L..45B,2005AN....326..849B,2005ragt.meet...39B} yields consistent results with the Continuum Fitting method for the spin of GRO J1655-40. This is important because assuming RPM in Kerr geometry the spin of GRO J1655-40 turns out to be much less than the predictions of Continuum Fitting/Fe-line method \cite{Motta:2013wga}. 

%%%%%%%%%%%%%%%%%%%%%%%%
%%%%%%%%%%%%%%%%%%%%%%%%
%%%%%%%%%%%%%%%%%%%%%%%%
%%%%%%%%%%%%%%%%%%%%%%%%
\begin{figure}[h!]
%\hspace{-1.8cm}
\subfloat[Non-axisymmetric Disk Oscillation Model 1  \label{Fig_6a}]{\includegraphics[scale=0.45]{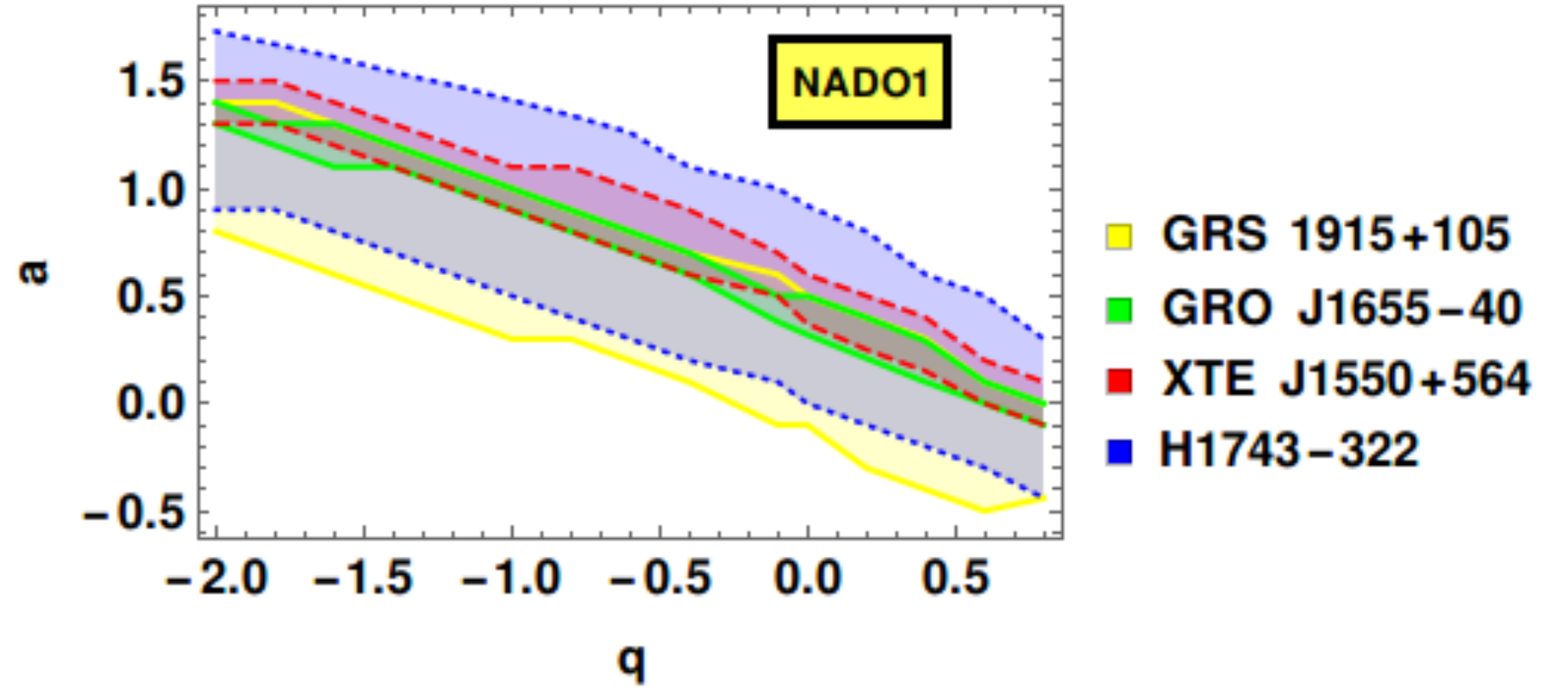}}
\subfloat[Non-axisymmetric Disk Oscillation Model 2 \label{Fig_6b}]{\includegraphics[scale=0.45]{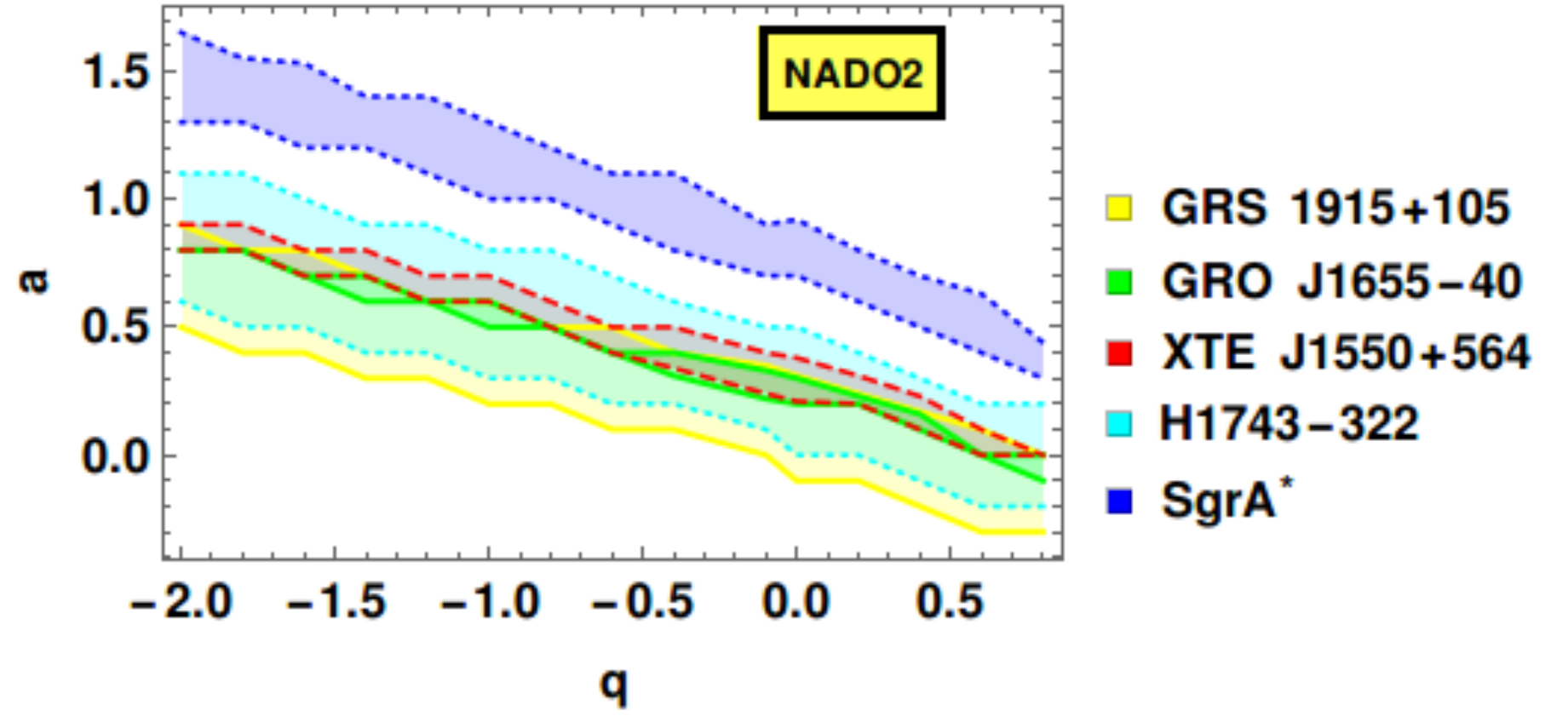}}
\caption{The above figure depicts the values of $q$ and $a$ that can reproduce the observed QPO frequencies for different BH sources (see \ref{Table1}) assuming the (a) Non-axisymmetric Disk Oscillation Model 1 (NADO1) and (b) Non-axisymmetric Disk Oscillation Model 2 (NADO2). See text for more discussions.}
\label{Fig_06}
\end{figure}
%%%%%%%%%%%%%%%%%%%%%%%%
%%%%%%%%%%%%%%%%%%%%%%%%
%%%%%%%%%%%%%%%%%%%%%%%%
%%%%%%%%%%%%%%%%%%%%%%%%

The second variant of the non-axisymmetric disk oscillation model (referred here as NADO2) assumes resonance between the $m=-1$ non-axisymmetric radial epicyclic frequency ($\nu_2=\nu_{per}=\nu_\phi-\nu_r$) with $m=-2$ non-axisymmetric vertical epicyclic frequency ($\nu_1=2\nu_\phi-\nu_\theta$) \cite{Torok:2010rk,Torok:2011qy,Kotrlova:2020pqy}. The resonances assumed in both NADO1 and NADO2 are expected to occur very close to the innermost stable cicular orbit, $r_{isco}$ \cite{Kotrlova:2020pqy,Torok:2011qy}. The resonant coupling assumed in NADO1 is considered unlikely as it involves coupling between an axisymmetric and a non-axisymmetric mode \cite{Horak:2008zg,Torok:2011qy}. The coupling between the pairs of oscillation modes considered in NADO 2 are allowed in principle, although the physical mechanism that would induce such a coupling is not yet well understood.

From \ref{Fig_6a} we note that for all the microquasars in \ref{Table1}, every $q$ value has some allowed values of $a$ which can fully explain the observed HFQPOs, assuming NADO1. Therefore, the sources GRS 1915+105, GRO J1655-40, XTE J1550+564 and H1743-322 trace a non-zero area in the $q-a$ plane of \ref{Fig_6a}. For SgrA* however, no value of $q$ can exactly reproduce its observed HFQPOs and hence it contributes maximally to the joint-$\chi^2$. When the second variant of the non-axisymmetric disk oscillation model (NADO2) is assumed, it is evident from \ref{Fig_6b} that all values of $q$ can explain the observed twin HFQPOs for every black hole source in \ref{Table1}.

\end{itemize} 
%%%%%%%%%%%%%%%%%%%%%%%%
%%%%%%%%%%%%%%%%%%%%%%%%
%%%%%%%%%%%%%%%%%%%%%%%%
%%%%%%%%%%%%%%%%%%%%%%%%

%%%%%%%%%%%%%%%%%%%%%%%%%%%%%%%%%%%%%%%%%%%%%%%%%%%%%%%%%%%%%%%%%%%%%%%%%%%%%%%%%%%%%%%%%%%%%%%%%%%
%%%%%%%%%%%%%%%%%%%%%%%%%%%%%%%%%%%%%%%%%%%%%%%%%%%%%%%%%%%%%%%%%%%%%%%%%%%%%%%%%%%%%%%%%%%%%%%%%%%
%%%%%%%%%%%%%%%%%%%%%%%%%%%%%%%%%%%%%%%%%%%%%%%%%%%%%%%%%%%%%%%%%%%%%%%%%%%%%%%%%%%%%%%%%%%%%%%%%%%
\section{Estimating the most favoured tidal charge parameter from the observed QPOs}\label{S5}

\begin{figure}[h!]
\centering
%\vspace{-3cm}
%\hspace*{-2cm}
\includegraphics[scale=0.58]{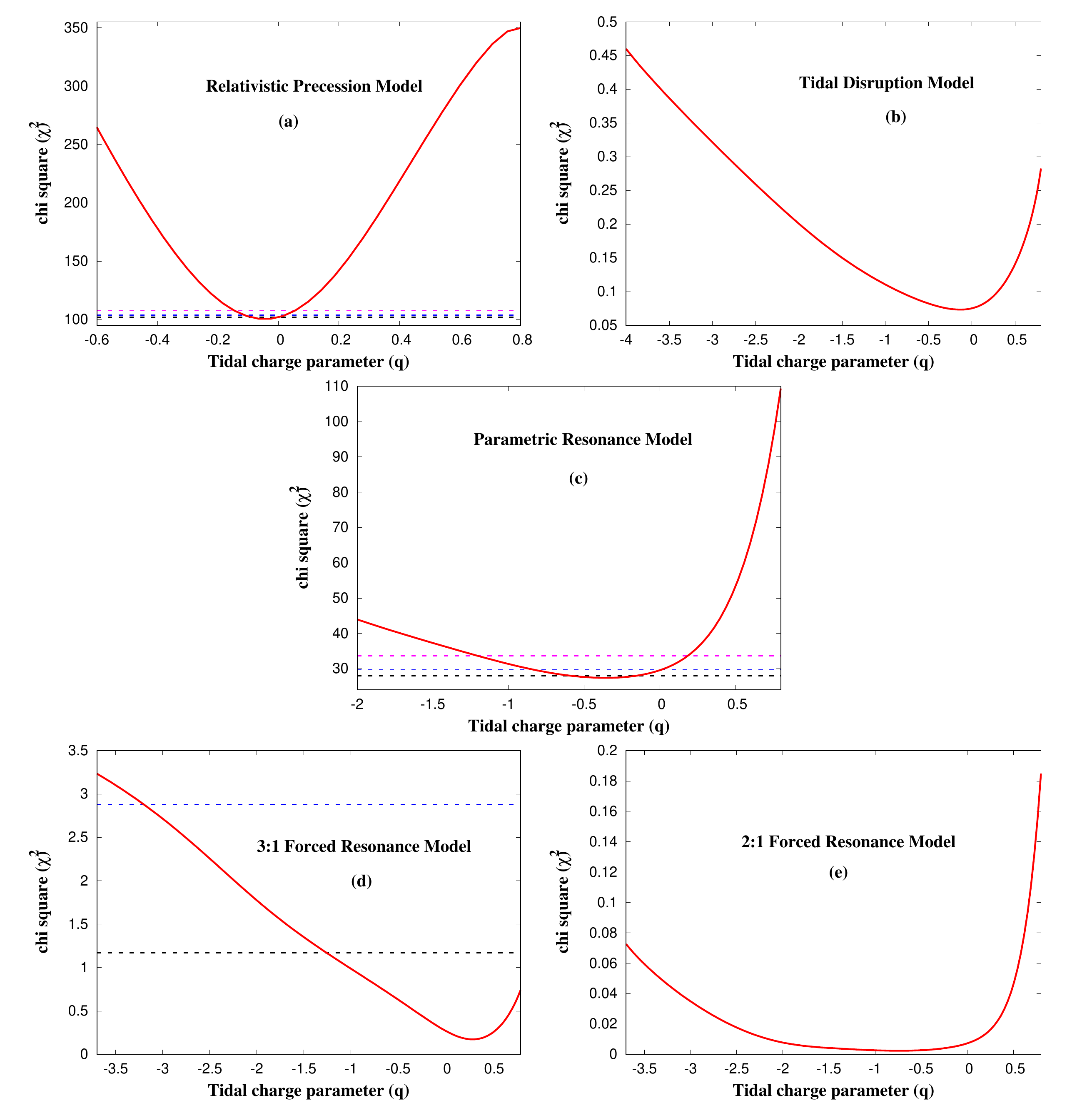}
\caption{The figure illustrates the variation of $\chi^{2}$ (evaluated using a sample of five black holes listed in \ref{Table1}) with the tidal charge parameter $q$ assuming --- (a) Relativistic Precession Model, (b) Tidal Disruption Model, (c) Parametric Resonance Model, (d) 3:1 Forced Resonance Model (FRM1) and (e) 2:1 Forced Resonance Model (FRM2). The minimum of $\chi^{2}$ denotes the most favoured tidal charge corresponding to different models. For more discussions one is referred to the main text.}
\label{Fig_07}
\end{figure}

\begin{figure}[h!]
\centering
%\vspace{-3cm}
%\hspace*{-2cm}
\includegraphics[scale=0.58]{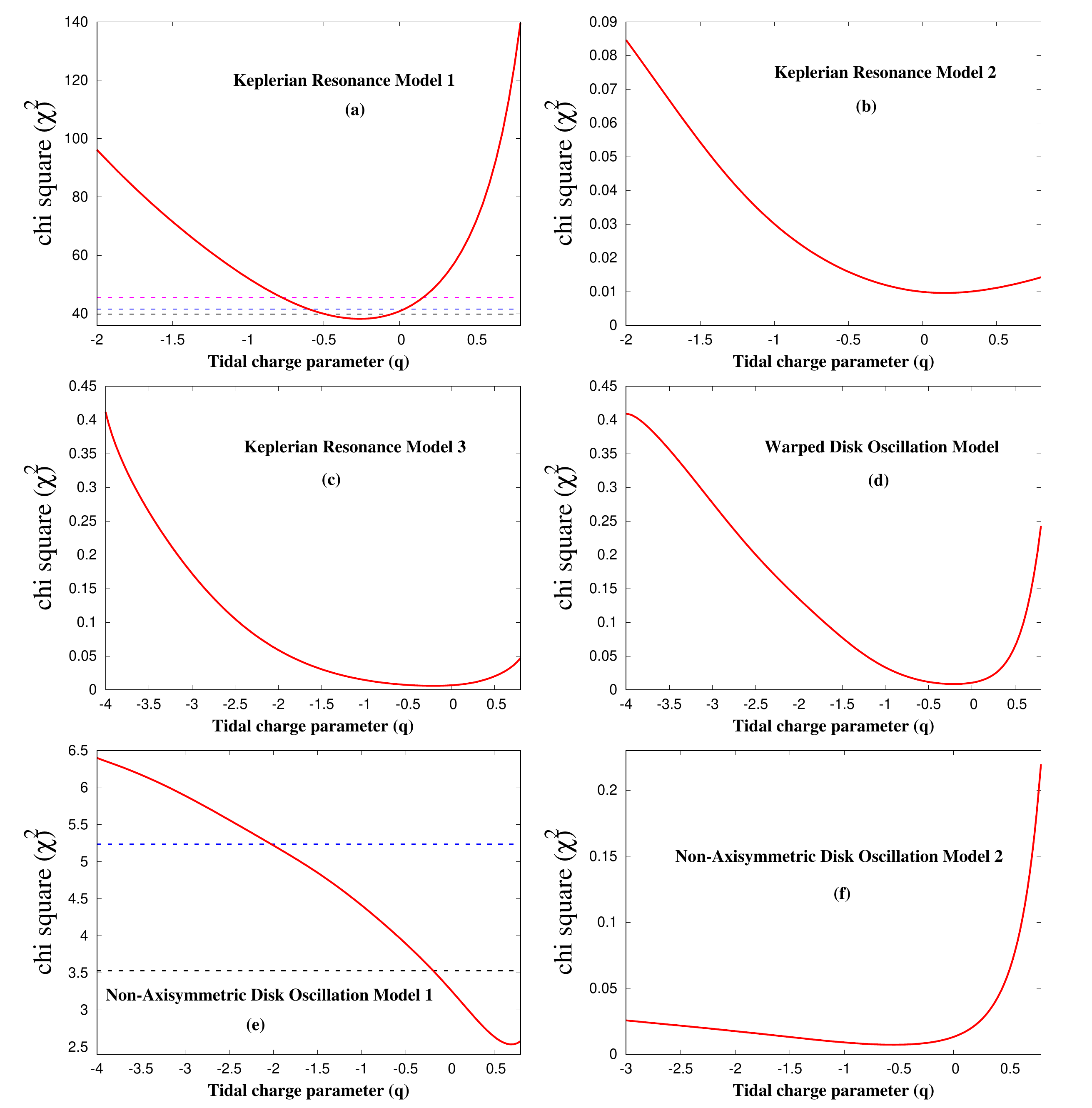}
\caption{The figure illustrates the variation of $\chi^{2}$ with the tidal charge parameter $q$ for a sample of five black holes listed in \ref{Table1}, assuming the following models: (a) Keplerian Resonance Model 1, (b) Keplerian Resonance Model 2, (c) Keplerian Resonance Model 3, (d) Warped Disc Oscillation Model, (e) Non-axisymmetric Disc Oscillation Model 1 and (f) Non-axisymmetric Disc Oscillation Model. The minimum of $\chi^{2}$ denotes the most favoured tidal charge corresponding to different models. For more discussions see the main text.}
\label{Fig_08}
\end{figure}

\begin{figure}[h!]
\centering
%\vspace{-3cm}
%\hspace*{-2cm}
\includegraphics[scale=0.72]{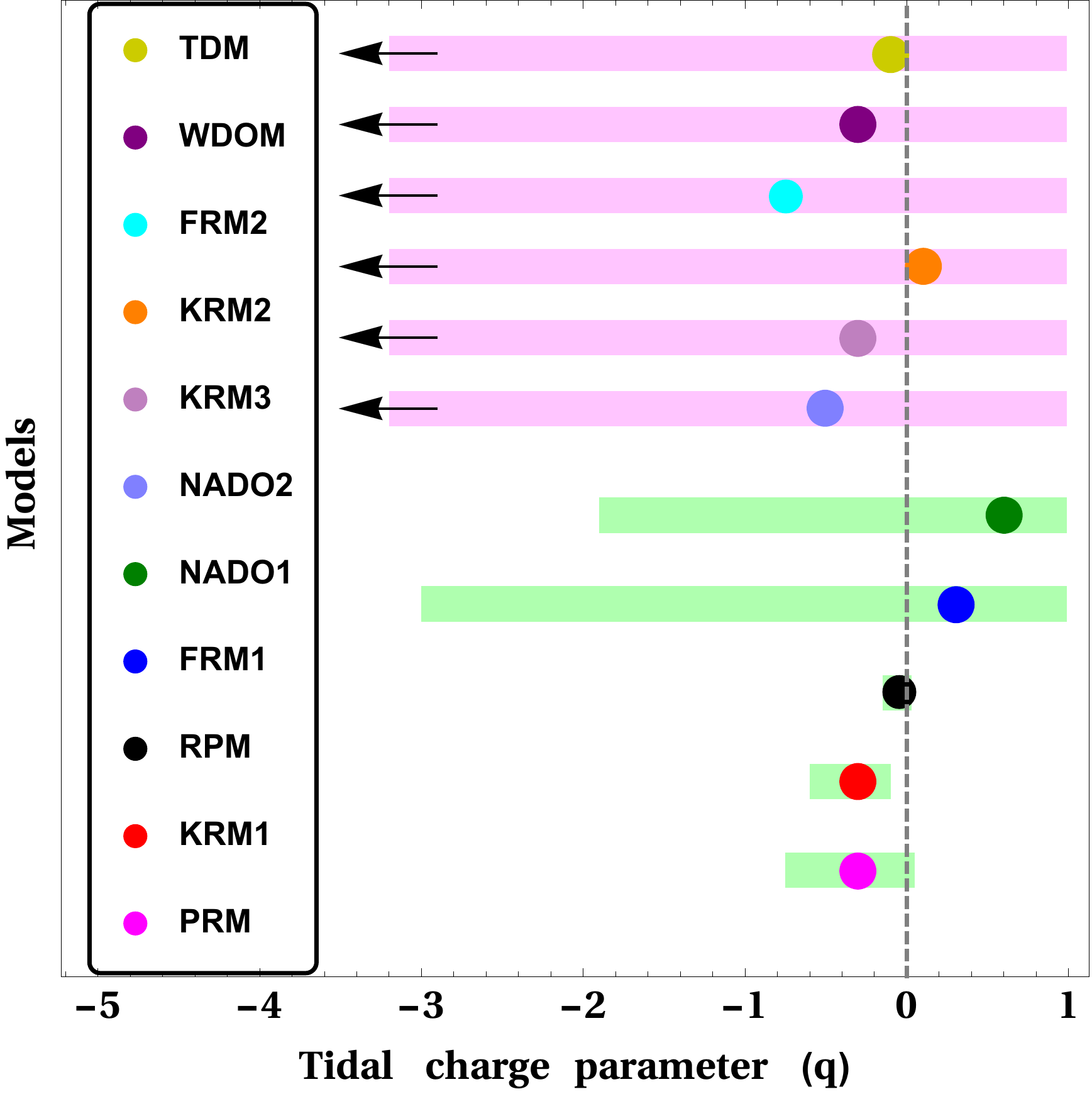}
\caption{The above figure summarizes the most favoured values of $q$ from each of the QPO models. For each of the models, the filled coloured circles represent the $q$ value corresponding to $\chi^2_{\rm min}$. The allowed values of $q$ from QPO observations within 90\% confidence interval for the models RPM, PRM, KRM1, FRM1 and NADO1 are shown with light green colored block. For the remaining six models the confidence intervals cannot be obtained. As a consequence they cover the entire range of x-axis, shaded with pink colored block. Since there's no theoretical restriction on the negative value of $q$, the leftmost part of the pink shaded regions are marked with arrows to indicate that they are in fact broader than the x-axis limits drawn in the figure. The maximum positive value of $q$ is unity and hence the pink shaded regions can extend maximum upto $q=1$ along the positive x-axis which is shown in the figure. The grey dashed line depicts \gr, i.e., $q=0$. As evident, among the 11 theoretical models, 8 of them predict negative values of $q$, while three of them predicts positive values. It is worth emphasizing that all of these QPO models are consistent with negative $q$ vales, within the 1-$\sigma$ confidence interval.}
\label{Fig_09}
\end{figure}

In the last section we have described the various kinematic and resonant models of QPOs proposed in the literature. We have also compared the theoretical QPO frequencies, predicted from each of these models, with the QPO frequencies observed in various BH sources listed in \ref{Table1}. Such a comparison enables us to understand the observationally allowed values of the tidal charge $q$ and rotation parameter $a$ in the purview of each of these models. In order to obtain a better understanding on this aspect we now perform an error analysis in this section. This requires computing the joint-$\chi^2$ as a function of $q$ for each of the models discussed in the last section. 
The $\chi ^{2}$ function, in the present context, is given by,
%%%%%%%%%%%%%%%%%%%%%%%%%%%%%%%%%%%%%%%%%%%%%%%%%%%%%%%%%%%%%%%%
\begin{align}
\label{S5-1}
\chi ^2 (q)=\sum_{i} \frac{\lbrace \nu_{\textrm{up1}{,i}}-\nu_1(q,a_{\rm min},M_{\rm min},r_{\rm min}) \rbrace ^2}{\sigma_{\nu_{\rm up1},i}^2}  
+ \sum_{i} \frac{\lbrace \nu_{\textrm{up2}{,i}}-\nu_2(q,a_{\rm min},M_{\rm min},r_{\rm min}) \rbrace ^2}{\sigma_{\nu_{\rm up2}, i}^2}~,
\end{align}
%%%%%%%%%%%%%%%%%%%%%%%%%%%%%%%%%%%%%%%%%%%%%%%%%%%%%%%%%%%%%%%%
which evaluates the difference between the observed and the theoretical model dependent QPO frequencies normalized by the observed errors in each of these frequencies. Here, $\nu_{\textrm{up1},i}$ and $\nu_{\textrm{up2},i}$ are respectively the observed upper and lower HFQPOs for the $i^{\rm th}$ source listed in \ref{Table1}, such that $i$ runs from $1$ to $5$. The observed errors in $\nu_{\textrm{up}1,i}$ and $\nu_{\textrm{up2},i}$ are respectively denoted by $\sigma_{\nu_{\rm up1}, i}$ and $\sigma_{\nu_{\rm up2}, i}$, which have been mentioned in \ref{Table1} as $\Delta\nu_{\rm up1}$ and $\Delta\nu_{\rm up2}$ for the $i^{\rm th}$ source. On the other hand, the theoretical, model dependent QPO frequencies are denoted by $\nu_1$ and $\nu_2$, respectively. For each of the QPO models, these theoretical frequencies have been summarized in \ref{Table2} and discussed in detail in \ref{S4}. It is clear from \ref{Table2} that the theoretical QPO frequencies $\nu_1$ and $\nu_2$ for a given model can be expressed in terms of $\nu_\phi$, $\nu_r$ and $\nu_\theta$, which in turn are functions of $q$, $a$, $r_{\rm em}$ and $M$. Following which, we adopt the procedure detailed below:

%%%%%%%%%%%%%%%%%%%%%%%%
%%%%%%%%%%%%%%%%%%%%%%%%
%%%%%%%%%%%%%%%%%%%%%%%%
%%%%%%%%%%%%%%%%%%%%%%%%
\begin{enumerate}

\item First of all, we select one of the theoretical models from \ref{Table2}, such that $\nu_1$ and $\nu_2$ are known functions of $q$, $a$, $r_{\rm em}$ and $M$.

\item Next we fix the tidal charge $q$ to some value, while satisfying the constraint: $q\leq 1$.

\item We then choose a BH source from \ref{Table1}, whose mass has been determined from other independent observations, as listed in \ref{Table1}. 

\item For the chosen value of the tidal charge $q$, we allow the spin to vary between $-\sqrt{1-q}\leq a \leq \sqrt{1-q}$, the emission radius $r_{\rm em}$ to vary between $r_{\rm ms}(q,a)\leq r_{\rm em} \leq r_{\rm ms}(q,a) + 20 R_{\rm g}$ and the mass $M$ of the chosen source to vary between $M_0-\Delta M \leq M \leq M_0 + \Delta M$ (where $M_0$ is the centroid value and $\Delta M$ is the error in mass given in \ref{Table1}). Given these variations in the respective quantities, we calculate $\chi^{2}$ for every combination of $a$, $M$ and $r_{\rm em}$. The values of $a$, $M$ and $r_{\rm em}$ that minimizes the $\chi^2$ (considering observations of both of the twin HFQPOs) are denoted by $a_{\rm min}$, $M_{\rm min}$ and $r_{\rm min}$ and are taken to be the observationally preferred mass, spin and excitation radius of the chosen source for the given $q$. This value of $\chi^2$ corresponding to $a_{\rm min}$, $M_{\rm min}$ and $r_{\rm min}$ for the $i^{th}$ source is denoted by $\chi^2_{m,i}$ which is essentially the $i^{th}$ term in the sum described by \ref{S5-1}. 

\item Keeping $q$ fixed, we repeat Step 3 and Step 4 upto the fifth source in \ref{Table1}, i.e., we note $\chi^2_{m,i}$ for all the sources and sum them up as in \ref{S5-1}. This gives us the value of $\chi^2(q)$ for the given $q$ with the model chosen in Step 1.

\item  We now repeat the steps 2 to 5 above, for a different value of $q$, keeping the QPO model fixed. This enables us to evaluate the variation of $\chi^2$ with $q$ for the chosen QPO model. The value of the tidal charge where $\chi^2$ minimizes (the minimum of $\chi^2$ being referred as $\chi^2_{\rm min}$) gives the observationally favoured value of $q$ (denoted by $q_{\rm min}$) within the domain of the chosen model. The variation of $\chi^2$ with $q$ for each of the QPO models is shown in \ref{Fig_07} and \ref{Fig_08}.

\item In case of models like RPM, which also addresses the observed low-frequency QPO, the form of $\chi^2$ is given by,
%%%%%%%%%%%%%%%%%%%%%%%%%%%%%%%%%%%%%%%%%%%%%%%%%%%%%%%%%%%%%%%%
\begin{align}
\label{S5-2}
\chi ^2 (q)&=\sum_{i} \frac{\lbrace \nu_{\textrm{up1}{,i}}-\nu_1(q,a_{\rm min},M_{\rm min},r_{\rm min}) \rbrace ^2}{\sigma_{\nu_{\rm up1}, i}^2}  
+\sum_{i}  \frac{\lbrace \nu_{\textrm{up2}{,i}}-\nu_2(q,a_{\rm min},M_{\rm min},r_{\rm min}) \rbrace ^2}{\sigma_{\nu_{\rm up2}, i}^2} 
\nonumber 
\\
&\hskip 4 cm +\frac{\lbrace \nu_{L,{\rm GRO}}-\nu_3(q,a_{\rm min},M_{\rm min},r_{\rm min}) \rbrace ^2}{\sigma_{\nu_{L},\rm GRO}^2}~.
\end{align}
%%%%%%%%%%%%%%%%%%%%%%%%%%%%%%%%%%%%%%%%%%%%%%%%%%%%%%%%%%%%%%%%
This takes into account the observed low-frequency QPO of GRO J1655-40 to arrive at the value of $q$, where $\chi^{2}$ minimizes. 

\item In \ref{Table3} and \ref{Table4} we report the values of $M_{\rm min}$ and $a_{\rm min}$ respectively for each of the BH sources corresponding to $q_{\rm min}$ (for values of $q_{\rm min}$ from different QPO models see \ref{Fig_07}-\ref{Fig_09}).
The values of $M_{\rm min}$ and $a_{\rm min}$ corresponding to $q\sim 0$ is also presented in these tables for completeness. These are then compared with the previously measured values of mass and spin for the BH sources.

\end{enumerate}  
%%%%%%%%%%%%%%%%%%%%%%%%
%%%%%%%%%%%%%%%%%%%%%%%%
%%%%%%%%%%%%%%%%%%%%%%%%
%%%%%%%%%%%%%%%%%%%%%%%%

%%%%%%%%%%%%%%%%%%%%%%%%
%%%%%%%%%%%%%%%%%%%%%%%%
%%%%%%%%%%%%%%%%%%%%%%%%
%%%%%%%%%%%%%%%%%%%%%%%%
\begin{table}[h!]
\vskip-1cm
\begin{center}
\hspace*{-2cm}
\begin{tabular}{|p{1.8cm}|p{2.8cm}|p{2.8cm}|p{3cm}|p{3cm}| p{3.5cm}| }
% \hline
% \multicolumn{5}{|c|}{Comparison of mass estimates} \\
\hline

$\rm Comparison$ & $\rm GRO ~J1655-40  $ & $\rm XTE ~J1550-564 $ & $\rm GRS ~1915+105 $ & $\rm H ~1743+322 $ & $\rm Sgr~A^*$\\
$\rm of ~mass $ & $\rm   $ & $\rm  $ & $\rm  $ & $\rm  $ & $\rm $\\
$\rm estimates$ & $\rm   $ & $\rm  $ & $\rm  $ & $\rm  $ & $\rm $\\
$\rm (\rm in ~M_\odot) $ & $\rm   $ & $\rm  $ & $\rm  $ & $\rm  $ & $\rm $\\
%\hline
\hline 
$\rm $ & $\rm $  & $\rm $ & $\rm  $ & $ \rm $ & $\rm $\\
$\rm Previous $ & $\rm 5.4\pm 0.3$ \cite{Beer:2001cg} & $ \rm 9.1\pm 0.61$ \cite{Orosz:2011ki}  & $ \rm 12.4^{+2.0}_{-1.8} $ \cite{Reid:2014ywa} & $ \rm 8.0-14.07$   & $ \rm (3.5-4.9) \rm \times 10^{-3}$\\ 
$\rm constraints$ & $ $ & $ $ & $ $ & $ $ \cite{Pei:2016kka,Bhattacharjee:2019vyy,Petri:2008jc}  & $  $ \cite{Ghez:2008ms,Gillessen:2008qv} \\

\hline
$\rm $ & $\rm $  & $\rm $ & $\rm  $ & $ \rm $ & $\rm $\\
$\rm RPM$ & $\rm 5.16 $  & $\rm  8.68$ & $\rm 13.26$ & $ \rm 9.77$ & $\rm 4.3\times 10^6$\\
$\rm $ & $\rm (q_{ \rm min}\sim -0.05)$  & $\rm (q_{ \rm min}\sim -0.05)$ & $\rm (q_{ \rm min}\sim -0.05)$ & $ \rm (q_{ \rm min}\sim -0.05)$ & $\rm (q_{ \rm min}\sim -0.05)$\\
$\rm $ & $\rm 5.24~(q\sim 0)$  & $\rm 8.94~ (q\sim 0)$ & $\rm 14.25~ (q\sim 0)$ & $ \rm 9.12~(q\sim 0)$ & $\rm 3.5\times 10^6~(q\sim 0)$\\

\hline
$\rm $ & $\rm $  & $\rm $ & $\rm  $ & $ \rm $ & $\rm $\\
%\vspace{0.2cm}
$\rm TDM$ & $\rm 5.68(q_{ \rm min}\sim -0.1)$  & $\rm 9.23(q_{ \rm min}\sim -0.1)$ & $\rm  12.57(q_{ \rm min}\sim-0.1)$ & $ \rm 11.17(q_{ \rm min}\sim-0.1)$ & $\rm 3.6\times 10^6(q_{ \rm min}\sim -0.1)$\\
$\rm $ & $\rm 5.33~(q\sim0)$  & $\rm 9.13~ (q\sim0)$ & $\rm 12.12 ~(q\sim0)$ & $ \rm 10.73~(q\sim0)$ & $\rm 3.5\times 10^6~(q\sim0)$\\
\hline 
$\rm $ & $\rm $  & $\rm $ & $\rm  $ & $ \rm $ & $\rm $\\
$\rm PRM$ & $\rm 5.44(q_{ \rm min}\sim -0.3)$  & $\rm 9.04(q_{ \rm min}\sim -0.3)$ & $\rm 12.0(q_{ \rm min}\sim -0.3)$ & $ \rm 11.56(q_{\rm min}\sim-0.3)$ & $\rm 3.5\times 10^6(q_{ \rm min}\sim -0.3)$\\
$\rm $ &  $5.16 \rm ~(q\sim0)$  & $\rm 9.64 ~(q\sim 0)$ & $\rm 14.37 ~(q\sim 0)$ & $ \rm 10.6~(q\sim 0)$ & $\rm 3.5\times 10^6~ (q\sim 0)$\\
\hline
$\rm $ & $\rm $  & $\rm $ & $\rm  $ & $ \rm $ & $\rm $\\
$\rm FRM1$ & $\rm 5.4(q_{ \rm min}\sim 0.3)$  & $\rm 8.91(q_{ \rm min}\sim 0.3)$ & $\rm  14.0(q_{ \rm min}\sim 0.3)$ & $ \rm 10.27(q_{ \rm min}\sim 0.3)$ & $\rm 3.5 \times 10^6 (q_{ \rm min}\sim 0.3)$\\
$\rm $ & $\rm 5.2~(q\sim 0)$  & $\rm  8.51~ (q\sim 0)$ & $\rm 13.2~ (q\sim 0)$ & $ \rm 12.57 ~(q\sim 0)$ & $\rm 3.5 \times 10^6 ~(q\sim 0)$\\
 \hline
 $\rm $ & $\rm $  & $\rm $ & $\rm  $ & $ \rm $ & $\rm $\\
 $\rm FRM2$ & $\rm 5.24 (q_{ \rm min}\sim -0.8)$  & $\rm  8.77(q_{ \rm min}\sim -0.8)$ & $\rm  12.76(q_{ \rm min}\sim -0.8)$ & $ 8.6 \rm (q_{ \rm min}\sim -0.8)$ & $\rm 3.5 \times 10^6 (q_{ \rm min}\sim -0.8)$\\
$\rm $ & $\rm 5.33 ~(q\sim 0)$  & $\rm  9.27~(q\sim 0)$ & $\rm  11.42~(q\sim 0)$ & $ \rm 9.28~(q\sim 0)$ & $\rm 3.6 \times 10^6~(q\sim 0)$\\
 \hline
$\rm $ & $\rm $  & $\rm $ & $\rm  $ & $ \rm $ & $\rm $\\
$\rm KRM1$ & $\rm 5.1 (q_{ \rm min}\sim -0.3)$  & $\rm  8.57(q_{ \rm min}\sim -0.3)$ & $\rm 12.21 (q_{ \rm min}\sim -0.3)$ & $ \rm 8.09(q_{ \rm min}\sim -0.3)$ & $\rm 3.5\times 10^6(q_{ \rm min}\sim -0.3)$\\
$\rm $ & $\rm 5.17(q\sim 0)$  & $\rm 8.49 (q\sim 0)$ & $\rm 12.13 (q\sim 0)$ & $ \rm 8.0(q\sim 0)$ & $\rm 3.5\times 10^6(q\sim 0)$\\
\hline 
$\rm $ & $\rm $  & $\rm $ & $\rm  $ & $ \rm $ & $\rm $\\
$\rm KRM2$ & $\rm 5.36 ~(q\sim 0)$  & $\rm 8.94 ~(q\sim 0)$ & $\rm 14.1 ~ (q\sim 0)$ & $ \rm 13.08 ~(q\sim 0)$ & $\rm 4.8 \times 10^6~(q\sim 0)$\\
\hline 
$\rm $ & $\rm $  & $\rm $ & $\rm  $ & $ \rm $ & $\rm $\\
$\rm KRM3$ & $\rm 5.7 (q_{ \rm min}\sim -0.3)$  & $\rm  9.23(q_{ \rm min}\sim -0.3)$ & $\rm  14.19(q_{ \rm min}\sim -0.3)$ & $ \rm 11.56(q_{ \rm min}\sim -0.3)$ & $\rm3.9 \times 10^6 (q_{ \rm min}\sim -0.3)$\\
$\rm $ & $5.67 \rm ~(q\sim 0)$  & $\rm 9.65 ~(q\sim 0)$ & $\rm 12.71~ (q\sim 0)$ & $ 10.33\rm ~(q\sim 0)$ & $\rm 4.2 \times 10^6~(q\sim 0)$\\
\hline
$\rm $ & $\rm $  & $\rm $ & $\rm  $ & $ \rm $ & $\rm $\\ 
$\rm WDOM$ & $\rm 5.66 (q_{ \rm min}\sim -0.3)$  & $\rm 8.55 (q_{ \rm min}\sim -0.3)$ & $\rm  13.26(q_{ \rm min}\sim -0.3)$ & $ \rm 10.5(q_{ \rm min}\sim -0.3)$ & $\rm 3.5\times 10^6 (q_{ \rm min}\sim -0.3)$\\
$\rm $ & $\rm 5.34 ~(q\sim 0)$  & $\rm 9.34 ~(q\sim 0)$ & $\rm 13.86~ (q\sim 0)$ & $ \rm 11.31 ~(q\sim 0)$ & $\rm 3.5 \times 10^6~ (q\sim 0)$\\
\hline 
$\rm $ & $\rm $  & $\rm $ & $\rm  $ & $ \rm $ & $\rm $\\
$\rm NADO1$ & $\rm 5.19(q_{ \rm min}\sim 0.6)$  & $\rm  8.78(q_{ \rm min}\sim 0.6)$ & $\rm 12.01 (q_{ \rm min}\sim 0.6)$ & $ \rm 8.46 (q_{ \rm min}\sim 0.6)$ & $\rm 3.5\times 10^6(q_{ \rm min}\sim 0.6)$\\
$\rm $ & $\rm 5.15~(q\sim 0)$  & $\rm 9.53 ~(q\sim 0)$ & $\rm  12.4~(q\sim 0)$ & $ \rm 11.13 ~(q\sim 0)$ & $\rm 3.5\times 10^6~ (q\sim 0)$\\
\hline
$\rm $ & $\rm $  & $\rm $ & $\rm  $ & $ \rm $ & $\rm $\\
$\rm NADO2$ & $\rm 5.67 (q_{ \rm min}\sim -0.5)$  & $\rm  8.96(q_{ \rm min}\sim -0.5)$ & $\rm  14.38(q_{ \rm min}\sim -0.5)$ & $ 10.0\rm (q_{ \rm min}\sim -0.5)$ & $\rm 4.8 \times 10^6 (q_{ \rm min}\sim -0.5)$\\
$\rm $ & $\rm 5.43~ (q\sim 0)$  & $\rm  9.22 ~(q\sim 0)$ & $\rm  14.04 ~(q\sim 0)$ & $ \rm 14.0~ (q\sim 0)$ & $\rm 4 \times 10^6~(q\sim 0)$\\
\hline\hline
%\multicolumn{5}{|c|}{PRM} \\

\end{tabular}
\caption{Comparison of mass estimates of the five BH sources obtained from each of the QPO models using $\chi^2$ minimization, with the previous estimates. For every model we report the preferred values of mass corresponding to $q \sim q_{\rm min}$ and also for $q\sim 0$ which corresponds to \gr.}
\label{Table3}
\end{center}

\end{table}
%%%%%%%%%%%%%%%%%%%%%%%%
%%%%%%%%%%%%%%%%%%%%%%%%
%%%%%%%%%%%%%%%%%%%%%%%%
%%%%%%%%%%%%%%%%%%%%%%%%

It is important to note that the previous estimates of mass are generally based on optical measurements which do not depend on the nature of strong gravity while earlier estimates of spin are sensitive to the near horizon geometry. Since the goal of the present work is to understand the viability of the braneworld model in explaining the QPO observations, we do not aim to derive the most favoured value of mass from QPO data but use the previous estimates of mass to arrive at our results. In this regard, we vary the mass of the sources in the range reported previously (see \ref{Table1}) and hence the most favoured value of mass from our analysis lies somewhere in the aforesaid range.

For spin however, we do not use the previously reported range as our guideline for two reasons: (a) The earlier estimates are based on GR, assuming Kerr geometry, while in our case the background metric differs from the Kerr spacetime; (b) Even after considering the Kerr background, different methods of constraining the spin (e.g., the Continuum Fitting method, Fe-line method, QPO observations) yield vastly different results for some black hole sources, e.g., GRO J1655-40 \cite{Motta:2013wga}. Therefore, we give another independent estimate of spin for the BH sources from QPO data considering the mass to be known. The spin corresponding to $q \sim q_{\rm min}$ (from each QPO model) is reported in \ref{Table4}, while the spin associated with $q\sim 0$ is also mentioned, so that it can be directly compared with previous estimates.

It is evident from \ref{Table4} that GRO J1655-40, GRS 1915+105 and Sgr A* exhibit a lot of disparity in the spin estimates from previous studies, all of which assume the background spacetime to be governed by the Kerr metric. For example, using the Continuum-Fitting method the spin of GRO J1655-40 turns out to be $0.65<a<0.75$ \cite{Shafee_2005}, Fe-line method yields $0.94<a<0.98$ \cite{Miller:2009cw}, while from QPO data one obtains $a=0.290\pm 0.003$ \cite{Motta:2013wga} (which is derived assuming RPM). From second column of \ref{Table4} we note that apart from PRM and KRM1 all the other QPO models predict a low value of spin for GRO J1655-40 as found in \cite{Motta:2013wga}. PRM and KRM1, on the other hand, predict high spin value for this source, which is in agreement with independent measurements based on Fe-line method \cite{Miller:2009cw}. 

For XTE J1550-564, the Continuum Fitting method yields a spin in the range $-0.11<a<0.71$, while the spin obtained from Fe-line method is $a=0.55^{+0.15}_{-0.22}$ \cite{Steiner:2010bt}. These are consistent with the results obtained from most of the QPO models, except PRM and KRM1, which once again predict high spin values for this source (third column of \ref{Table4}). The previous spin estimates of the source GRS 1915+105 again yields quite discrepant results, since from Continuum Fitting itself the spin turns out to be maximal $a\sim 0.98$ \cite{McClintock:2006xd} as well as intermediate $a\sim 0.7$ \cite{2006MNRAS.373.1004M}. The Fe-line method however, implies $a \sim 0.6- 0.98$ \cite{Blum:2009ez}. Using revised mass and inclination for this source, a more recent study based on Continuum Fitting method \cite{Mills:2021dxs} report a larger range for the spin of the source, i.e., $0.4<a<0.98$. Our results based on PRM and KRM1 (Column 4 of \ref{Table4}) provide the best agreement with previous studies \cite{McClintock:2006xd,Blum:2009ez,Mills:2021dxs}. Other QPO models (Column 4 of \ref{Table4}), on the other hand, predict lower values of spin. 

Using the Continuum-Fitting method the spin of H1743-322 is constrained to be $a=0.2\pm 0.3 $ (with 68\% confidence) while, one also obtains the following bound for the spin parameter: $a<0.92$, for this source (with 99.7\% confidence) \cite{Steiner:2011kd}. This is mostly in agreement with our results (Column five of \ref{Table4}), except for PRM and KRM1, which seem to generically predict a high spin value for all the sources considered here. Sgr A* does not exhibit any prominent X-ray reflection features. This stems from its low accretion rate, which makes the accretion flow hot, tenuous and optically-thin to X-ray emission \cite{Rees:1982pe,Narayan:1995ic}. Its spin is therefore inferred from GRMHD simulations, which are used to predict its radio spectrum and frequency-dependent polarization as a function of system parameters \cite{Reynolds:2013rva,Moscibrodzka:2009gw,Shcherbakov:2010ki}. Due to the uncertainties associated in computing its complex radio continuum, constraints on its spin exhibit a lot of discrepancy, e.g. $a\sim 0.9$ \cite{Moscibrodzka:2009gw}, $a\sim 0.5$ \cite{Shcherbakov:2010ki}. Analysis of the X-ray light curve of Sgr A* indicates a maximal spin for this object ($ \rm a=0.9959\pm 0.0005$) \cite{2010MmSAI..81..319A}, while another study based on the motion of S2 stars reveal that $a\lesssim 0.1$ \cite{Fragione:2020khu} for this source. From the present analysis we find that the source exhibits high/near maximal spin from all the QPO models (sixth column of \ref{Table4}), which in turn is in agreement with \cite{Moscibrodzka:2009gw,2010MmSAI..81..319A}.

%%%%%%%%%%%%%%%%%%%%%%%%
%%%%%%%%%%%%%%%%%%%%%%%%
%%%%%%%%%%%%%%%%%%%%%%%%
%%%%%%%%%%%%%%%%%%%%%%%%
\begin{table}[h!]
\vskip-1.8cm
\begin{center}
\hspace*{-2cm}
\begin{tabular}{|p{1.8cm} |p{3cm}|p{3.5cm}|p{3cm}|p{2.8cm}| p{3.3cm}| }
% \hline
% \multicolumn{5}{|c|}{Comparison of mass estimates} \\
\hline

$\rm Comparison$ & $\rm GRO ~J1655-40  $ & $\rm XTE ~J1550-564 $ & $\rm GRS ~1915+105 $ & $\rm H ~1743+322 $ & $\rm Sgr~A^*$\\
$\rm of ~spin $ & $\rm   $ & $\rm  $ & $\rm  $ & $\rm  $ & $\rm $\\
$ \rm estimates $ & $\rm   $ & $\rm  $ & $\rm  $ & $\rm  $ & $\rm $\\
$ \rm  $ & $\rm   $ & $\rm  $ & $\rm  $ & $\rm  $ & $\rm $\\
%\hline
\hline 
$\rm Previous$ & $\rm a\sim 0.65-0.75 $ \cite{Shafee_2005} & $ \rm -0.11<a<0.71$ \cite{Steiner:2010bt}  & $ a\sim \rm 0.98 $ \cite{McClintock:2006xd} & $ a=\rm 0.2 \pm 0.3 $ \cite{Steiner:2011kd} & $ \rm a\sim 0.92$ \cite{Moscibrodzka:2009gw}\\ 
$\rm constraints$ & $ \rm a\sim 0.94-0.98$ \cite{Miller:2009cw} &  $ $ & $ a\sim \rm 0.7 $ \cite{2006MNRAS.373.1004M} & $ \rm $   & $ \rm a\sim 0.5 $ \cite{Shcherbakov:2010ki} \\
$\rm $ & $ \rm a=0.29\pm 0.003  $ \cite{Motta:2013wga} & $ \rm a= 0.55^{+0.15}_{-0.22}$ \cite{Steiner:2010bt}  & $ \rm a\sim 0.6-0.98 $ \cite{Blum:2009ez} &   $   $   & $ \rm a=0.9959\pm 0.0005$ \cite{2010MmSAI..81..319A}\\ 
$\rm $ & $ $  & $ $ & $ \rm a\sim 0.4-0.98$ \cite{Mills:2021dxs} &  $  $   & $ \rm a \sim 0.1 $ \cite{Fragione:2020khu} \\
\hline
$\rm $ & $\rm $  & $\rm $ & $\rm   $ & $ \rm $ & $\rm $\\ 
$\rm RPM$ & $\rm 0.3 $  & $\rm 0.37 $ & $\rm  0.28 $ & $ \rm 0.33$ & $\rm 1.04 $\\
$\rm $ & $\rm (q_{ \rm min}\sim -0.05)$  & $\rm (q_{ \rm min}\sim -0.05)$ & $\rm (q_{ \rm min}\sim -0.05)$ & $ \rm (q_{ \rm min}\sim -0.05)$ & $\rm (q_{ \rm min}\sim -0.05)$\\
$\rm $ & $\rm 0.29~(q\sim 0)$  & $\rm 0.36~(q\sim 0)$ & $\rm 0.32~ (q\sim 0)$ & $ \rm  0.21~(q\sim 0)$ & $\rm 0.92~ (q\sim 0)$\\
\hline
$\rm $ & $\rm $  & $\rm  $ & $\rm  $ & $ \rm$ & $\rm $\\
$\rm TDM$ & $\rm 0.23(q_{ \rm min}\sim -0.1)$  & $\rm 0.29(q_{ \rm min}\sim -0.1)$ & $\rm  0(q_{ \rm min}\sim -0.1)$ & $ \rm 0.29(q_{ \rm min}\sim -0.1)$ & $\rm  1.04(q_{ \rm min}\sim -0.1)$\\
$\rm $ & $\rm 0.1~(q\sim 0)$  & $\rm 0.23~(q\sim 0)$ & $\rm -0.1~ (q\sim 0)$ & $ \rm  0.2~(q\sim 0)$ & $\rm 0.99~ (q\sim 0)$\\
\hline
$\rm $ & $\rm $  & $\rm  $ & $\rm  $ & $\rm  $ & $\rm  $ \\
$\rm PRM$ & $\rm 1.1(q_{ \rm min}\sim -0.3)$  & $\rm 1.1(q_{ \rm min}\sim -0.3)$ & $\rm 1.0 (q_{ \rm min}\sim -0.3)$ & $ \rm 1.18 (q_{ \rm min}\sim -0.3)$ & $\rm 1.18 (q_{ \rm min}\sim -0.3)$\\
$\rm $ & $\rm 0.9~(q\sim 0)$  & $\rm 0.95~(q\sim 0)$ & $\rm 0.92~ (q\sim 0)$ & $ \rm  0.97~(q\sim 0)$ & $\rm 0.99~ (q\sim 0)$\\
\hline
$\rm $ & $\rm $ & $\rm $ & $\rm $ & $ \rm $   &  $\rm   $\\
$\rm FRM1$ & $\rm 0.25(q_{ \rm min}\sim 0.3)$  & $\rm 0.30(q_{ \rm min}\sim 0.3)$ & $\rm 0.24 (q_{ \rm min}\sim 0.3)$ & $ \rm 0.29 (q_{ \rm min}\sim 0.3)$ & $\rm  0.89(q_{ \rm min}\sim 0.3)$\\
$\rm $ & $\rm 0.3~(q\sim 0)$  & $\rm 0.34~(q\sim 0)$ & $\rm 0.26~ (q\sim 0)$ & $ \rm  0.6~(q\sim 0)$ & $\rm 0.97~ (q\sim 0)$\\
\hline
 $\rm $ & $\rm  $ & $\rm  $ & $ \rm $ & $ $ &  $ $ \\
$\rm FRM2$ & $\rm 0.4(q_{ \rm min}\sim -0.8)$  & $\rm 0.5(q_{ \rm min}\sim -0.8)$ & $\rm  0.3(q_{ \rm min}\sim -0.8)$ & $ \rm 0.2(q_{ \rm min}\sim -0.8)$ & $\rm  1.3(q_{ \rm min}\sim -0.8)$\\
$\rm $ & $\rm 0.1~(q\sim 0)$  & $\rm 0.25~(q\sim 0)$ & $\rm -0.2~ (q\sim 0)$ & $ \rm 0.0 ~(q\sim 0)$ & $\rm 0.97~ (q\sim 0)$\\
 \hline
$\rm $ & $\rm  $  & $\rm  $ & $\rm  $ & $\rm  $  &  $ $ \\
$\rm KRM1$ & $\rm 1.18(q_{ \rm min}\sim -0.3)$  & $\rm 1.18(q_{ \rm min}\sim -0.3)$ & $\rm 1.18 (q_{ \rm min}\sim -0.3)$ & $ \rm 1.18 (q_{ \rm min}\sim -0.3)$ & $\rm  1.18 (q_{ \rm min}\sim -0.3)$\\
$\rm $ & $\rm 0.99~(q\sim 0)$  & $\rm 0.99~(q\sim 0)$ & $\rm 0.97~ (q\sim 0)$ & $ \rm  0.99~(q\sim 0)$ & $\rm 0.99~ (q\sim 0)$\\
\hline 
$\rm $ & $\rm  $  & $\rm  $ & $\rm  $ & $\rm  $  &  $ $ \\
%$\rm KRM2$ & $\rm (q\sim -0.4)$  & $\rm (q\sim -0.4)$ & $\rm  (q\sim -0.4)$ & $ \rm (q\sim -0.4)$ & $\rm  (q\sim -0.4)$\\
$\rm KRM2$ & $\rm 0.32~(q\sim 0)$  & $\rm 0.36~(q\sim 0)$ & $\rm 0.32~ (q\sim 0)$ & $ \rm  0.6~(q\sim 0)$ & $\rm 0.97~ (q\sim 0)$\\
\hline 
$\rm $ & $\rm  $  & $\rm $ & $\rm  $ & $\rm  $  &  $ $\\
$\rm KRM3$ & $\rm 0.27 (q_{ \rm min}\sim -0.3)$  & $\rm 0.29(q_{ \rm min}\sim -0.3)$ & $\rm  0.20(q_{ \rm min}\sim -0.3)$ & $ \rm 0.4(q_{ \rm min}\sim -0.3)$ & $\rm 1.0 (q_{ \rm min}\sim -0.3)$\\
$\rm $ & $\rm 0.22~(q\sim 0)$  & $\rm 0.3~(q\sim 0)$ & $\rm 0.0~ (q\sim 0)$ & $ \rm  0.22~(q\sim 0)$ & $\rm 0.99~ (q\sim 0)$\\
\hline 
$\rm $ & $\rm $ & $\rm  $ & $\rm  $ & $\rm  $  &  $ \rm  $\\
$\rm WDOM$ & $\rm 0.35(q_{ \rm min}\sim -0.3)$  & $\rm 0.31(q_{ \rm min}\sim -0.3)$ & $\rm  0.20(q_{ \rm min}\sim -0.3)$ & $ \rm 0.33(q_{ \rm min}\sim -0.3)$ & $\rm  1.18(q_{ \rm min}\sim -0.3)$\\
$\rm $ & $\rm 0.1~(q\sim 0)$  & $\rm 0.26~(q\sim 0)$ & $\rm 0.1~ (q\sim 0)$ & $ \rm  0.26~(q\sim 0)$ & $\rm 0.99~ (q\sim 0)$\\
\hline 
$\rm $ & $\rm $  & $\rm  $ & $\rm  $ & $\rm  $  &  $\rm  $ \\
$\rm NADO1$ & $\rm 0.0 (q_{ \rm min}\sim 0.6)$  & $\rm 0.1(q_{ \rm min}\sim 0.6)$ & $\rm  -0.2(q_{ \rm min}\sim 0.6)$ & $ \rm -0.2(q_{ \rm min}\sim 0.6)$ & $\rm  0.63(q_{ \rm min}\sim 0.6)$\\
$\rm $ & $\rm 0.36~(q\sim 0)$  & $\rm 0.6~(q\sim 0)$ & $\rm 0.22~ (q\sim 0)$ & $ \rm  0.6~(q\sim 0)$ & $\rm 0.99~ (q\sim 0)$\\
\hline 
$\rm $ & $\rm $  & $\rm  $ & $\rm  $ & $\rm  $   &  $\rm  $ \\
$\rm NADO2$ & $\rm 0.32 (q_{ \rm min}\sim -0.5)$  & $\rm 0.32 (q_{ \rm min}\sim -0.5)$ & $\rm  0.29(q_{ \rm min}\sim -0.5)$ & $ \rm 0.28(q_{ \rm min}\sim -0.5)$ & $\rm  0.9(q_{ \rm min}\sim -0.5)$\\
$\rm $ & $\rm 0.25~(q\sim0)$  & $\rm 0.31~(q\sim 0)$ & $\rm 0.24~ (q\sim 0)$ & $ \rm 0.5 ~(q\sim 0)$ & $\rm 0.8~ (q\sim 0)$\\
\hline

 %\multicolumn{5}{|c|}{PRM} \\
 \hline

\end{tabular}
\caption{Comparison of spin estimates of the five BH sources obtained from each of the QPO models using $\chi^2$ minimization, with the previous estimates. For every model we report the preferred values of spin corresponding to $q\sim q_{\rm min}$ and also for $q\sim 0$ which corresponds to \gr.}
\label{Table4}
\end{center}

\end{table}
%%%%%%%%%%%%%%%%%%%%%%%%
%%%%%%%%%%%%%%%%%%%%%%%%
%%%%%%%%%%%%%%%%%%%%%%%%
%%%%%%%%%%%%%%%%%%%%%%%%

It is to be emphasized that, though the theoretical QPO frequencies $\nu_{1}$ and $\nu_{2}$ are actually functions of 4 parameters, namely, $q$, $a$, $M$ and $r_{\rm em}$, but we minimize $\chi^{2}$ only with respect to the tidal charge $q$. This stems from the fact that our goal in this work is to discern the most favoured value of $q$ from the QPO observations in black holes. This implies that all the four parameters are not on equal footing, i.e., they can be classified into two categories: (i) the ``interesting parameters", whose values are determined simultaneously from $\chi^2$ minimization and (ii) the ``uninteresting parameters", whose values are estimated by keeping the ``interesting parameters" fixed to different values and minimizing the $\chi^2$ \cite{1976ApJ...210..642A}. In our case, the tidal charge $q$ is the ``interesting parameter", while the ``uninteresting parameters" are $a$, $M$ and $r_{\rm em}$. Depending on the number of ``interesting parameters" the confidence intervals, or, $\Delta \chi^{2}$ from $\chi^{2}_{\rm min}$ are determined. For example, in the present case, $\Delta \chi^2$ corresponding to 68\%, 90\% and 99\% confidence intervals are numerically identical to 1, 2.71 and 6.63 \cite{1976ApJ...210..642A}, respectively. It is important to note that, while performing this analysis, we keep the tidal charge $q$ fixed for all the five black holes, while we vary $M$, $a$ and $r_{\rm em}$. This is because the tidal charge is related to the electric part of the bulk Weyl tensor and epitomizes the signature of the extra dimensions on the brane. In our analysis we assume that the extra dimension affects all the braneworld black holes uniformly, which justifies our method of analysis \cite{Banerjee:2017hzw,Banerjee:2019sae}.

%%%%%%%%%%%%%%%%%%%%%%%%
%%%%%%%%%%%%%%%%%%%%%%%%
%%%%%%%%%%%%%%%%%%%%%%%%
%%%%%%%%%%%%%%%%%%%%%%%%
%%%%%%%%%%%%%%%%%%%%%%%%
%%%%%%%%%%%%%%%%%%%%%%%%
%%%%%%%%%%%%%%%%%%%%%%%%
%%%%%%%%%%%%%%%%%%%%%%%%

As discussed above, the error estimator $\chi^{2}$ is a function of the tidal charge parameter $q$ and hence we depict its functional dependence in   \ref{Fig_07} and \ref{Fig_08} for various QPO models, which forms the basis of this work. \ref{Fig_07} illustrates the variation of $\chi^2$ with the tidal charge $q$ assuming --- (a) Relativistic Precession Model (RPM), (b) Tidal Disruption Model (TDM), (c) Parametric Resonance Model (PRM), (d) 3:1 Forced Resonance Model (FRM1) and (e) 2:1 Forced Resonance Model (FRM2). While in \ref{Fig_08} we have plotted the variation of $\chi^2$ with $q$ for (a) Keplerian Resonance Model 1 (KRM1), (b) Keplerian Resonance Model 2 (KRM2), (c) Keplerian Resonance Model 3 (KRM3) (d) Warped Disc Oscillation Model (WDOM) (e) Non-axisymmetric Disc Oscillation Model 1 (NADO1) and (f) Non-axisymmetric Disc Oscillation Model 2 (NADO2). 
In each of the sub-figures within \ref{Fig_07} and \ref{Fig_08} the value of the tidal charge $q$, where $\chi^{2}$ attains its minimum is the one favoured by observations. This value of $q$ is denoted by $q_{\rm min}$. The $1-\sigma$, $2-\sigma$ and $3-\sigma$ confidence intervals have also been plotted for the QPO models, wherever possible, in \ref{Fig_07} and \ref{Fig_08} with black, blue and magenta dashed lines, respectively. These correspond to $\chi^2=\chi^2_{\rm min}+1$, $\chi^2=\chi^2_{\rm min}+2.71$ and $\chi^2=\chi^2_{\rm min}+6.63$, respectively.  
 
From \ref{Fig_07} we note that $\chi^2$ attains a minimum for a \emph{negative} tidal charge for the following models --- (i) Relativistic Precession Model ($q_{\rm min}\simeq -0.05$), (ii) Tidal Disruption Model ($q_{\rm min}\simeq -0.1$), (iii) Parametric Resonance Model ($q_{\rm min}\simeq -0.3$) and (iv) 2:1 Forced Resonance Model ($q_{\rm min}\simeq -0.8$). However for the 3:1 Forced Resonance Model, positive value of the tidal charge parameter seems to be more favoured. Similarly in \ref{Fig_08}, $\chi^2$ minimizes for a \emph{negative} $q$ when --- (i) Keplerian Resonance Model 1 ($q_{\rm min}\simeq -0.3$), (ii) Keplerian Resonance Model 3 ($q_{\rm min}\simeq -0.3$) (iii) Warped Disc Oscillation Model ($q_{\rm min}\simeq -0.3$) and (iv) Non-axisymmetric Disc Oscillation Model 2 ($q_{\rm min}\simeq -0.5$) are considered. On the other hand, the Keplerian Resonance Model 2 and the Non-axisymmetric Disc Oscillation Model 1 predicts minimum $\chi^{2}$ for a positive value of the tidal charge parameter $q$. 

We have emphasized before, that a negative tidal charge is clearly a non-GR signature, which can be realized in a higher dimensional scenario and our result shows that eight out of the eleven QPO models considered here favour a \emph{negative} $q$. For the remaining models, a positive $q$ is favoured, which can be realized both in GR and also in the extra dimensional scenario considered here. Thus we see that the existence of an additional spatial dimension is favoured, if we want to explain the observed QPO frequencies using various theoretical models in the context of braneworld black holes. 

From \ref{Fig_07} we note that the two most widely used QPO models, namely, the Relativistic Precession Model (RPM) and the Parametric Resonance Model (PRM) not only favour a \emph{negative} tidal charge $q$, but also rules out \gr\ ($q=0$) within $1-\sigma$. However, when the $2-\sigma$ interval is considered both these models include the Kerr scenario. For RPM, the values of $q$ allowed within $3-\sigma$ confidence level lie in the interval $-0.15\leq q \leq 0.03$, while for PRM this interval corresponds to $-1.2\leq q \leq 0.2$. The results obtained from KRM1 (see \ref{Fig_08}) are also in accordance with the findings from RPM and PRM, since this model also rules out GR within $1-\sigma$. In particular, $\chi^2_{\rm min}$ is attained at $q_{\rm min}\simeq -0.3$, while the tidal charge within $-0.75\leq q\leq 0.2$ is in the $3-\sigma$ interval. Confidence intervals can also be drawn for FRM1 (see \ref{Fig_07}) and NADO1 (see \ref{Fig_08}). Both of these models favour a positive tidal charge with $\chi^{2}$ attaining minimum at $q\simeq 0.3$ and $q\simeq 0.6$ for FRM1 and NADO1 respectively. However, these models span a large range of $q$ within $90\%$ confidence level, viz, $-3.2\lesssim q\lesssim 0.99$ (FRM1) and $-1.9\lesssim q \lesssim 0.99$ (NADO1). The results based on RPM, PRM and FRM1 are however on a firmer ground compared to NADO1 or KRM1 because the resonant couplings assumed in the last two models are difficult to realize in the accretion scenario (see \ref{Sec4-2}). We however consider them for the sake of completeness. 
 
The resonant models like PRM can more naturally explain the commensurability of QPO frequencies while RPM is also a widely used QPO model. Both PRM and RPM not only favor a \emph{negative} tidal charge but also establish stringent constraints on the same, in particular, they rule out very high positive or negative values of $q$, consistent with previous studies \cite{Bhattacharya:2016naa,Banerjee:2019sae}. The other models favouring a positive $q$ (also realizable in the higher dimensional setup) allows a wide range of the tidal charge within $2-\sigma$ and hence cannot provide any strong bound on the tidal charge. The other models, where confidence contours cannot be drawn, cannot impose strong constraints on the tidal charge.

%%%%%%%%%%%%%%%%%%%%%%%%
%%%%%%%%%%%%%%%%%%%%%%%%
%%%%%%%%%%%%%%%%%%%%%%%%
%%%%%%%%%%%%%%%%%%%%%%%%
%%%%%%%%%%%%%%%%%%%%%%%%
%%%%%%%%%%%%%%%%%%%%%%%%
%%%%%%%%%%%%%%%%%%%%%%%%
%%%%%%%%%%%%%%%%%%%%%%%%

%%%%%%%%%%%%%%%%%%%%%%%%
%%%%%%%%%%%%%%%%%%%%%%%%
%%%%%%%%%%%%%%%%%%%%%%%%
%%%%%%%%%%%%%%%%%%%%%%%%
%%%%%%%%%%%%%%%%%%%%%%%%
%%%%%%%%%%%%%%%%%%%%%%%%
%%%%%%%%%%%%%%%%%%%%%%%%
%%%%%%%%%%%%%%%%%%%%%%%%

From the above discussion along with \ref{Fig_07} and \ref{Fig_08} we note that the observations related to QPOs in black holes generically favour a \emph{negative} tidal charge parameter. This is in accordance with some of our earlier findings based on the image of M87* \cite{Banerjee:2019nnj}, as well as on the continuum spectrum of black holes \cite{Banerjee:2019cjk,Banerjee:2019sae}. It is interesting to note that independent observations, e.g., continuum spectrum, QPOs, black hole shadow based on different black hole samples consistently favoured negative values of the tidal charge parameter.  Finally, to summarize, our results derived from all the QPO models have been presented in \ref{Fig_09}, where we report the value of $q_{\rm min}$ from each of the models by coloured filled circles along with confidence intervals, wherever possible (90\% for RPM, PRM, KRM1, FRM1 and NADO1). For the remaining six models the confidence intervals cannot be obtained. As a consequence, they cover the entire range of x-axis, shaded with pink colored block. Since there's no theoretical restriction on the negative value of $q$, the leftmost part of the pink shaded regions are marked with arrows to indicate that they are in fact broader than the x-axis limits drawn in the figure. The maximum positive value of $q$ is unity and hence the pink shaded regions can extend maximum upto $q=1$ along the positive x-axis, which is shown in the figure.
%%%%%%%%%%%%%%%%%%%%%%%%
%%%%%%%%%%%%%%%%%%%%%%%%
%%%%%%%%%%%%%%%%%%%%%%%%
%%%%%%%%%%%%%%%%%%%%%%%%

%%%%%%%%%%%%%%%%%%%%%%%%
%%%%%%%%%%%%%%%%%%%%%%%%
%%%%%%%%%%%%%%%%%%%%%%%%
%%%%%%%%%%%%%%%%%%%%%%%%

%%%%%%%%%%%%%%%%%%%%%%%%
%%%%%%%%%%%%%%%%%%%%%%%%
%%%%%%%%%%%%%%%%%%%%%%%%
%%%%%%%%%%%%%%%%%%%%%%%%

%%%%%%%%%%%%%%%%%%%%%%%%%%%%%%%%%%%%%%%%%%%%%%%%%%%%%%%%%%%%%%%%%%%%%%%%%%%%%%%%%%%%%%%%%%%%%%%%%%%
%%%%%%%%%%%%%%%%%%%%%%%%%%%%%%%%%%%%%%%%%%%%%%%%%%%%%%%%%%%%%%%%%%%%%%%%%%%%%%%%%%%%%%%%%%%%%%%%%%%
%%%%%%%%%%%%%%%%%%%%%%%%%%%%%%%%%%%%%%%%%%%%%%%%%%%%%%%%%%%%%%%%%%%%%%%%%%%%%%%%%%%%%%%%%%%%%%%%%%%
\section{Concluding Remarks}\label{S6}

In this work we explore the signatures of extra dimensions from the quasi-periodic oscillations observed in the power density spectrum of black holes. We consider a well studied and motivated higher dimensional scenario, where our visible universe is confined in a 3-brane embedded in a five dimensional bulk spacetime governed by Einstein gravity. The presence of an extra spatial dimension modifies the gravitational field equations on the brane by the introduction of projected bulk Weyl tensor and quadratic terms in the brane energy-momentum tensor. The stationary and axisymmetric vacuum black hole solution on the brane, resembles the Kerr-Newmann metric in GR, but differs as the tidal charge carried by the braneworld black holes, inherited from the bulk Weyl tensor, can assume \emph{negative} values. This turns out to be a distinctive signature of the higher dimensions, which we wish to observe, if present, from the QPO frequencies associated with black holes. It is worthwhile to emphasize that the QPOs provide a cleaner probe to the background metric compared to the continuum spectrum or the iron line. Since these QPOs originate from local or collective motion of accreting plasma near the marginally stable circular orbit of black holes and are solely dependent on the background metric and not on the complex and ill-understood processes of the accretion flow. Thus QPOs can potentially be a more effective probe of the background spacetime, which we have explored in detail in this work. 

Previous effort in this direction has attempted to extract the signature of the tidal charge from the observed QPO frequencies of two sources, namely GRS 1915+105 and Sgr A*, and reported that a negative tidal charge can explain the observations better, when the parametric resonance model is assumed \cite{Stuchlik:2008fy}. The present work, on the other hand,  is a more comprehensive and complete endeavour in this direction, where we have not only considered nearly all the available observations of HFQPOs in black holes (the four micro-quasars, viz., GRO J1655-40, GRS 1915+105, XTE J1550+564, H 1743+322 and the galactic centre supermassive black hole Sgr A*) but have also compared them with the theoretical predictions from the kinematic and the resonant QPO models existing in the literature. Such a study has enabled us to decipher the observationally favoured values of $q$ within the domain of each of these models. In order to establish stronger constraints on the tidal charge we further perform a $\chi^{2}$ analysis by comparing the theoretical QPO frequencies from each of the models described in \ref{Table2}, with the observations from five sources. Our analysis reveals that out of the eleven QPO models considered in this work, eight models, namely, RPM, TDM, PRM, FRM2, KRM1, KRM3, WDOM and NADO2, favour a \emph{negative} tidal charge, a distinguishing signature of extra dimensions. In particular, the Relativistic Precession Model and the Parametric Resonance Model, the two most extensively used models of QPOs, not only favour a negative $q$, but also exclude \gr\ within $1-\sigma$ confidence interval. When the $2-\sigma$ interval is considered, such models do include the Kerr scenario, although they consistently rule out large positive or negative values of the tidal charge within $3-\sigma$, a result in agreement with earlier findings \cite{Bhattacharya:2016naa,Banerjee:2019sae}. Although similar constraints on $q$ can also be established from the Keplerian Resonance Model 1 (KRM1), the resonant coupling between the orbital and the radial epicyclic motion is difficult to realize in this model within the purview of accretion scenario. The remaining five models, namely TDM, FRM2, KRM3, WDOM and NADO2, exhibiting minimum $\chi^2$ for a negative tidal charge, do not provide any strong constraints on $q$, unlike RPM and PRM models. Further, the three models (FRM1, KRM2 and NADO1), where the $\chi^{2}$ minimizes for a positive $q$, does not disfavour the extra dimensional scenario.

While it might seem from the above discussion that the observed QPO frequencies in the four micro-quasars (GRO J1655-40, GRS 1915+105, XTE J1550+564, H 1743+322) and Sgr A* indicate a preference towards braneworld black holes, compared to the Kerr scenario, it must be noted that these results are based on the various pre-existing QPO models. In this regard it may be noted that despite observing HFQPOs for decades there is no clear indication on the choice of the correct theoretical model, if any. There has been general consensus on the fact that HFQPOs should not be caused by different physics in different models, but rather a single model, that is yet to be determined, should explain all the observations. Thus, even though majority of the QPO models prefer a negative value for the tidal charge, it is immature to claim that braneworld scenario is more favoured compared to \gr. As and when a more complete QPO model is arrived at, one can perform the analysis presented here, to see whether braneworld scenario is indeed preferred or not. This works provides the framework necessary for such an analysis. At this outset we must also admit that the present analysis has certain limitations arising from --- (i) uncertainties associated with the physical mechanism giving rise to the QPOs, (ii) the small observational sample (a larger sample can potentially impose much more stringent constraints on $q$) and (iii) the lack of precise data. The precision of the data is expected to improve by more than an order of magnitude with the launch of the ESA (European Space Agency) X-ray mission LOFT (Large Observatory for X-ray Timing), which will enhance the scope to explore the QPO phenomenology in near future.

%%%%%%%%%%%%%%%%%%%%%%%%%%%%%%%%%%%%%%%%%%%%%%%%%%%%%%%%%%%%%%%%%%%%%%%%%%%%%%%%%%%%%%%%%%%%%%%%%%%
%%%%%%%%%%%%%%%%%%%%%%%%%%%%%%%%%%%%%%%%%%%%%%%%%%%%%%%%%%%%%%%%%%%%%%%%%%%%%%%%%%%%%%%%%%%%%%%%%%%
%%%%%%%%%%%%%%%%%%%%%%%%%%%%%%%%%%%%%%%%%%%%%%%%%%%%%%%%%%%%%%%%%%%%%%%%%%%%%%%%%%%%%%%%%%%%%%%%%%%
\section*{Acknowledgements}

The research of SSG is partially supported by the Science and Engineering Research Board-Extra Mural Research Grant No. (EMR/2017/001372), Government of India. Research of S.C. is funded by the INSPIRE Faculty Fellowship (Reg. No. DST/INSPIRE/04/2018/000893) from the Department of Science and Technology, Government of India and by the Start-Up Research Grant from SERB, DST, Government of India (Reg. No. SRG/2020/000409).

%%%%%%%%%%%%%%%%%%%%%%%%%%%%%%%%%%%%%%%%%%%%%%%%%%%%%%%%%%%%%%%%%%%%%%%%%%%%%%%%%%%%%%%%%%%%%%%%%%%
%%%%%%%%%%%%%%%%%%%%%%%%%%%%%%%%%%%%%%%%%%%%%%%%%%%%%%%%%%%%%%%%%%%%%%%%%%%%%%%%%%%%%%%%%%%%%%%%%%%
%%%%%%%%%%%%%%%%%%%%%%%%%%%%%%%%%%%%%%%%%%%%%%%%%%%%%%%%%%%%%%%%%%%%%%%%%%%%%%%%%%%%%%%%%%%%%%%%%%%
%Bibliography
\bibliography{new-ref,KN-ED,QPO,Brane,IB,Gravity_3_partial,Gravity_1_full,Black_Hole_Shadow}

\bibliographystyle{./utphys1}
%%%%%%%%%%%%%%%%%%%%%%%%%%%%%%%%%%%%%%%%%%%%%%%%%%%%%%%%%%%%%%%%%%%%%%%%%%%%%%%%%%%%%%%%%%%%%%%%%%%
%%%%%%%%%%%%%%%%%%%%%%%%%%%%%%%%%%%%%%%%%%%%%%%%%%%%%%%%%%%%%%%%%%%%%%%%%%%%%%%%%%%%%%%%%%%%%%%%%%%
%%%%%%%%%%%%%%%%%%%%%%%%%%%%%%%%%%%%%%%%%%%%%%%%%%%%%%%%%%%%%%%%%%%%%%%%%%%%%%%%%%%%%%%%%%%%%%%%%%%
\end{document}